\pdfoutput=1

\documentclass{jfm}
\usepackage{graphicx}
\usepackage{newtxtext}
\usepackage[slantedGreek]{newtxmath}
\usepackage{comment}
\usepackage[linesnumbered,ruled]{algorithm2e}
\usepackage{upgreek}
\usepackage{hyperref}
\usepackage{wrapfig}
\newsavebox{\tempfig}
\usepackage{graphicx, floatrow}
\usepackage[utf8]{inputenc}
\usepackage[english]{babel}
\usepackage{amsmath}
\usepackage{amssymb}
\usepackage{bm}
\usepackage[utf8]{inputenc}
\usepackage[english]{babel}
\usepackage[table,usenames,dvipsnames]{xcolor}
 \usepackage{textcomp}
 \usepackage[table]{xcolor}
 \usepackage{multirow}
\usepackage[square,numbers]{natbib}
\usepackage{tikz}
\usetikzlibrary{plotmarks}
\newcommand{\be}{\begin{equation}}
\newcommand{\ee}{\end{equation}}
\usepackage{graphicx}
\usepackage{etoolbox}
\usepackage{comment}
\usepackage{booktabs}
\DeclareRobustCommand\full  {\tikz[baseline=-0.6ex]\draw[blue,thick] (0,0)--(0.5,0);}
\DeclareRobustCommand\fullgreen{\tikz[baseline=-0.6ex]\draw[green!60!black,thick] (0,0)--(0.5,0);}
\DeclareRobustCommand\fullblack  {\tikz[baseline=-0.6ex]\draw[black,thick] (0,0)--(0.5,0);}
\DeclareRobustCommand\fullblackthick  {\tikz[baseline=-0.6ex]\draw[black,very thick] (0,0)--(0.5,0);}
\DeclareRobustCommand\fullred  {\tikz[baseline=-0.6ex]\draw[red,thick] (0,0)--(0.5,0);}
\DeclareRobustCommand\fullredthick  {\tikz[baseline=-0.6ex]\draw[red,very thick] (0,0)--(0.5,0);}

\DeclareRobustCommand\fullorange  {\tikz[baseline=-0.6ex]\draw[orange,thick] (0,0)--(0.5,0);}

\DeclareRobustCommand\dashedgreen{\tikz[baseline=-0.6ex]\draw[green!60!black,thick,dashed] (0,0)--(0.54,0);}

\DeclareRobustCommand\dashedred{\tikz[baseline=-0.6ex]\draw[red,thick,dashed] (0,0)--(0.54,0);}
\DeclareRobustCommand\dashedblack{\tikz[baseline=-0.6ex]\draw[black,thick,dashed] (0,0)--(0.54,0);}

\newrobustcmd*{\mycircle}[1]{\tikz{\filldraw[draw=#1,fill=#1] (0,0) circle [radius=0.08cm];}}
\newrobustcmd*{\mycirclesmall}[1]{\tikz{\filldraw[draw=#1,fill=#1] (0,0) circle [radius=0.02cm];}}
\newrobustcmd*{\mytriangle}[1]{\tikz{\filldraw[draw=#1,fill=#1] (0,0)--(0.2cm,0) -- (0.1cm,0.2cm);}}
\newrobustcmd*{\mysquare}[1]{\tikz{\filldraw[draw=#1,fill=#1] (0,0)--(0.15cm,0) -- (0.15cm,0.15cm) -- (0,0.15cm);}}
\makeatletter
\renewcommand\@makefnmark{\hbox{\@textsuperscript{\normalfont\@thefnmark}}}
\makeatother
\def\del#1{\textcolor{black}{}}
\def\rm#1{\textcolor{black}{#1}}
\def\rv#1{\textcolor{black}{#1}} 

\shorttitle{Ship as a buoy}
\shortauthor{Y. Du, G. Wang and Y. Pan}

\title{Ship-as-a-buoy for Phase-resolved Reconstruction of Surface Waves}

\author{Yifan Du\aff{1}\corresp{\email{yifandu@umich.edu}}, Guangyao Wang \aff{1}
 \and Yulin Pan\aff{1}}

\affiliation{\aff{1} University of Michigan, Ann Arbor}

\begin{document}

\maketitle

\begin{abstract}
We develop a data assimilation framework to address the ``ship-as-a-buoy'' problem, i.e., estimation of the wave field from time series of ship motion. Unlike previous algorithms which only provide information of wave spectrum, our approach reconstructs the instantaneous phase-resolved wave field. The full framework includes two steps: the first step to generate an initial (rough) estimation of the wave field using the response-amplitude-operators (RAOs), and the second step to refine the reconstruction using an ensemble Kalman filter (EnKF) to incorporate the measured motion time series. Together with EnKF in the second step, we model the coupled wave-ship dynamics by a high-order spectral (HOS) method for waves and a Cummins equation for ship motion (although higher-fidelity ship model can be applied). We conduct extensive tests of our method for the case with one ship, and the naturally extended cases with multiple ships or with uncertain parameters in the ship motion model. For all test cases we demonstrate favorable reconstruction of the wave field, which in general further improves with the number of ships available and with uncertain parameters properly estimated. We also provide insights on the successful reconstruction in terms of ensemble correlations between the motion time series and features of the wave field.   

\end{abstract}

\vspace*{12pt}
\begin{keywords}
Ocean surface wave, data assimilation, ship as a buoy
\end{keywords}


\section{Introduction}\label{sec: introduction}


The ``ship as a buoy’’ problem refers to the estimation of sea state using measured ship motions. Traditionally, the problem is usually solved under necessary assumptions to obtain only phase-averaged quantities such as wave spectra. 
\rv{In this paper, we address a phase-resolved extension of the ``ship as a buoy’’ problem. Specifically, we develop a fundamentally new approach to enable the instantaneous reconstruction of phase-resolved ocean surface through time series of ship motions, via data assimilation. }
The developed approach is flexible to be extended to applications with multiple ships and works well even when some parameters in the ship motion model are uncertain.

The ``ship as a buoy’’ problem is itself considered difficult even if only the wave spectrum is desired in the traditional manner \citep{nielsen2023estimating}. The underlying reason is that the problem is in nature an under-determined problem, with the degree of freedom in a directional spectrum far larger than that in the time series of body motions (e.g., a frequency component in the motion can be induced by waves from different directions). Most existing approaches are based on applications of frequency-domain transfer functions, known as Response Amplitude Operators (RAOs) \citep{tannuri2003estimating, pascoal2008non, nielsen2008wave, pascoal2017estimation}. 
To overcome the under-determined nature of the problem, i.e., a high-dimensional linear system with infinitely many possible solutions of directional wave spectrum, \citet{iseki2000bayesian} propose to include an additional regularization term penalizing the non-smoothness of the wave spectrum. The problem thus turns into an optimization problem which seeks one solution of directional wave spectrum that is sufficiently “smooth” and “approximately” generates the body response \citep{ren2021sea}. However, this optimization problem requires significant computational cost due to its high-dimensional nature and the need to estimate a hyper-parameter measuring the strength of the regularization term (which is addressed through a Bayesian information criterion in \citet{nielsen2006estimations, nielsen2008introducing}). 
\rv{Some machine learning based methods \citep{scholcz2020ship,han2022directional} are also developed, yet their accuracy is only guaranteed for the combinations of sea state and ship configurations that are covered in the training data \citep{nielsen2024comparing}. Therefore, we focus on the physics based models that are proven to be most generalizable across different wave and ship configurations. }

\rv{While the RAO-based spectral estimation methods have been successfully implemented and the results are to some extent verified in experiments \citep{stredulinsky2010quest,nielsen2012sea}, their limitations are obvious. The most important limitation for this line of works is the lack of phase information in the estimated sea states. It remains unknown whether a phase-resolved reconstruction of the wave field is possible solely from the motion measurements of a floating ship. Although the RAO itself contains information about the phase lag between wave excitation and ship response \citep{fossen2011handbook,newman2018marine}, such phase information is in general not sufficiently explored in the reconstruction of the wave field. There are some preliminary trials of phase-resolved reconstruction within the RAO-based framework\citep{takami2022reconstruction, takami2024onboard}, yet these works only attempt to reconstruct a time sequence of incidence wave at midship location rather than a full spatiotemporal wave field.}

\rv{Even if only the wave spectrum is desired, the RAO-based methods have clear limitations.} First, the accuracy of these methods is limited by linear assumptions in both the wave and ship motion models, with the obtained spectrum assumed (but not necessarily) stationary in an artificially chosen time window \citep{nielsen2017concise}. Second, while these methods are in principle extendable to applications with multiple ships \citep{mounet2022simultaneous,nielsen2019sea}, the existing extensions are based on a simple weighted superposition of the results from different ships, rather than a unified framework incorporating all ship measurements. Third, the RAOs describing ship motions are assumed to be known and fixed, which may however not be the case for ships at sea \citep{fossen2011handbook} (e.g., added mass can vary depending on ship displacement condition \citep{sepehrirahnama2019fluid} and is not easy to recalculate instantaneously). 

\rv{In this paper, we aim to solve an extended version of the ``ship-as-a-buoy'' problem, to reconstruct instantaneous phase-resolved ocean waves from measurements of floating body motions. 
For this purpose, we develop a fundamentally new approach in time domain that}
overcomes all limitations summarized above. Our approach consists of three components: (1) a high-order spectral (HOS) method \citep{dommermuth1987high} to solve the Euler equation of nonlinear wave evolution; (2) a Cummins-equation-based model to compute ship motion in a given wave field (assuming long-wave approximation); and (3) an ensemble-based data assimilation interface which provides optimal correction to the state (wave and motion) according to the measured motion. In addition, we develop a parameter estimation component based on the maximum-likelihood method, which can be applied when there are uncertain parameters involved in the ship model. 

Regarding (3), data assimilation technique has been widely studied in fluid mechanics in recent years \citep{mons2021ensemble,du2023state} as a tool to infuse measurement data into numerical models. In general, there are two types of strategies in data assimilation, namely smoothing and filtering \citep{law2015data}, with the former seeking a model evolution that best matches the measurements in a fixed time window \citep{lions1971optimal}, and the latter improving the modeled state by adding an optimal correction derived from measurement data \citep{evensen1994sequential,evensen2003ensemble}. Both types of strategies have been applied in conjunction to the HOS method with radar/buoy data (i.e., not in ``ship as a buoy’’ context), the former by \citet{wang2021phase} and the latter by \citet{aragh2008variation,wu2022improved}. We continue to explore the ensemble-based filtering approach, which better meets the need of real-time wave forecast as discussed in \citet{wang2022phase}. Compared to our previous works \citep{wang2021phase} and \citep{wang2022phase}, assimilation of ship motion instead of wave elevation imposes another layer of complexity (since wave elevation can now only be indirectly inferred from its correlation with ship motion), which we address in the current paper. 

Another difficulty involved in the present framework is the lack of an initial condition for the wave field to start the ensemble-based data assimilation, since no wave measurement is available. Such an initial condition therefore needs to be estimated through the ship motion data, which we develop a new RAO-based method to address. 
Our approach removes the under-determinedness of the problem through an assumption in dominating wave traveling direction (to be discussed later in \S \ref{subsec:initial guess from RAO}), which in the meanwhile turns the high-dimensional linear problem into a low-dimensional nonlinear one that is solved through quasi-Newton method with much lower computational cost compared to \citet{nielsen2006estimations}. As we will show, the initial condition prescribed in such a way is accurate enough for the subsequent data assimilation process. 


The detailed formulation of our method, as outlined above, is presented in Section 2 of the paper. In Section 3, we carry out extensive verification of the method \rv{on a small floating body that models a typical small- to medium-sized unmanned ocean surface vehicle. This case is slightly simpler than those with large ships considered in the classical ``ship as a buoy'' works, but contains the essential physical elements of the problem}. We first show that with one ship present, our method can reconstruct the wave field with an error of one order of magnitude lower than the initial condition, whereas the initial condition estimated by our RAO-based method can be already comparable in its accuracy to previous methods \citep{takami2024onboard,nielsen2006estimations, iseki2000bayesian}, if implemented in a phase-resolved manner. We then provide results with multiple ships present, and show that error of reconstruction can be further significantly reduced with the increase of ship number. We last demonstrate the application of our method in cases where some parameters in the ship model (e.g., certain elements in the added mass matrix in the Cummins equation) are uncertain. In addition to successful reconstruction of the ocean surface, our method also gives an estimation of these parameters that converge to their true values. We conclude the paper in Section 4.

\section{Theoretical formulation and computational approach}
\label{sec:method}
\begin{figure}
    \centering  \includegraphics[width=1.0\textwidth]{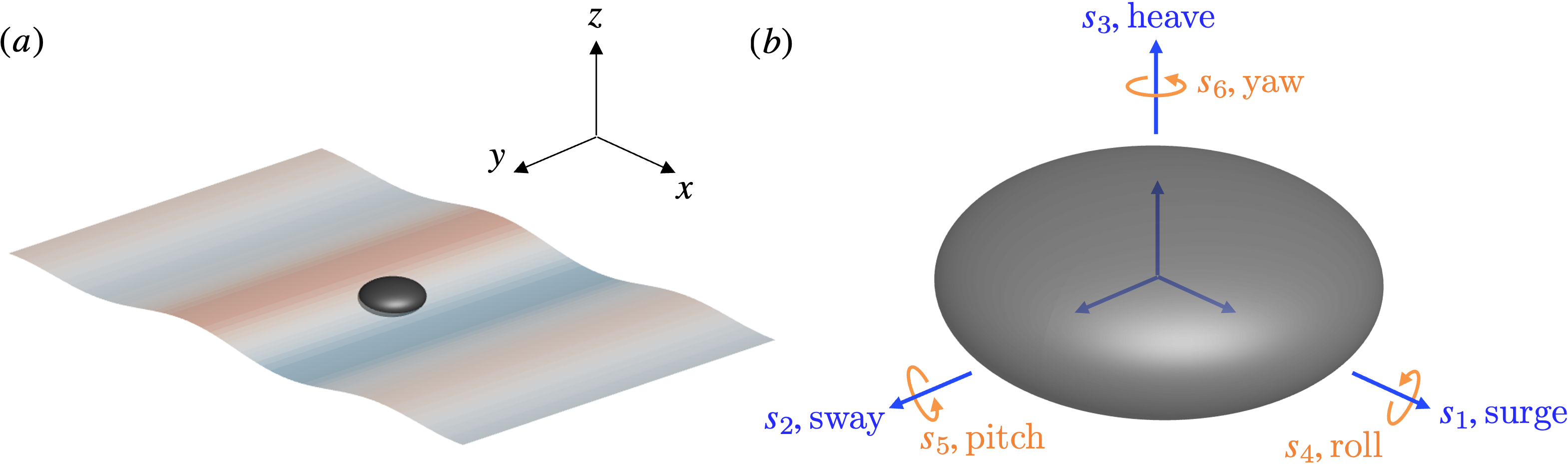}
    \caption{Schematics of the wave-body interaction: $(a)$ An ellipsoid floating on the surface of a wave field; $(b)$ Definition of six degrees of freedom $s_1$ to $s_6$ of the body.}
    \label{fig:wave body schm}
\end{figure}
\begin{figure}
    \centering  \includegraphics[width=1.0\textwidth]{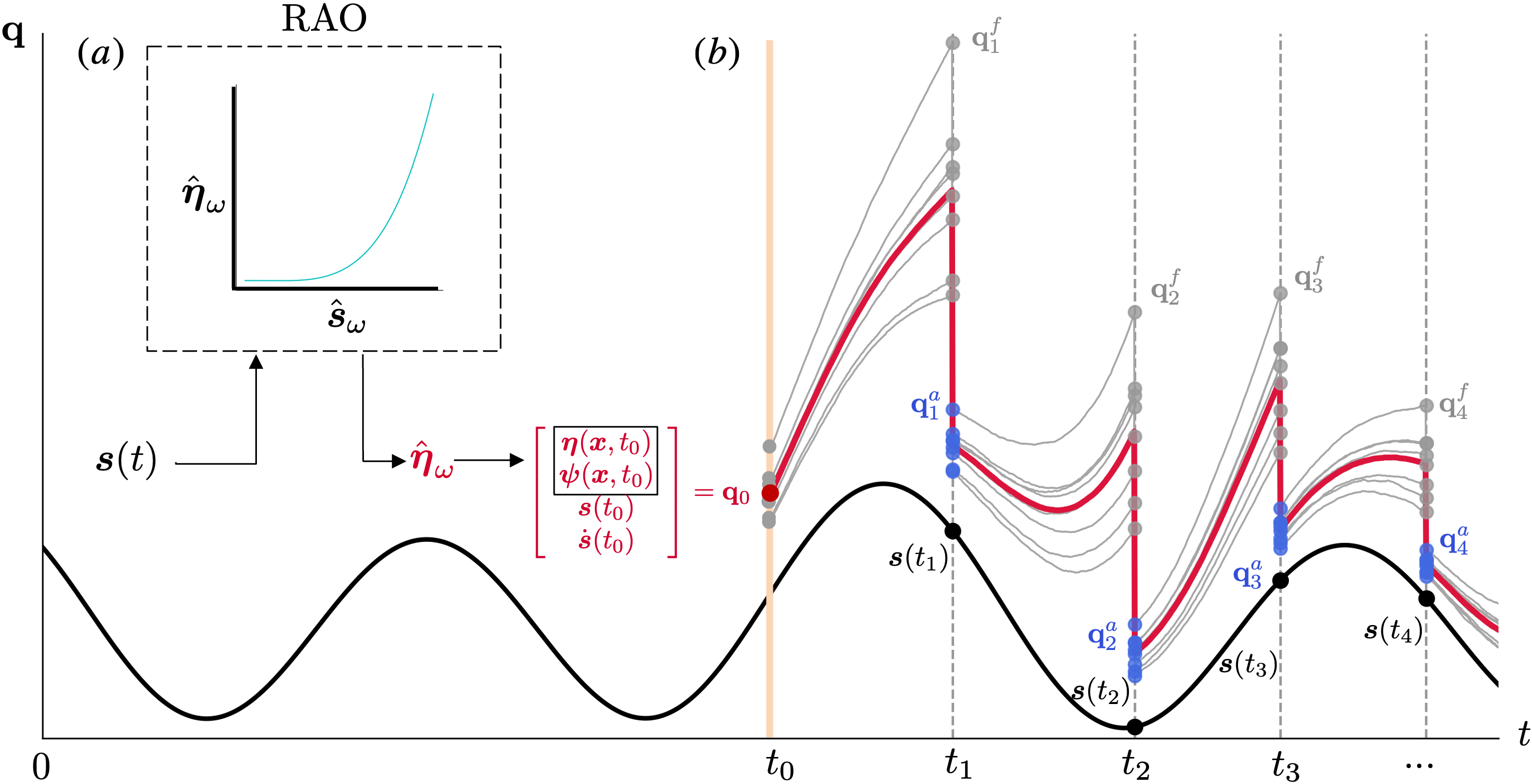}
    \caption{Schematics of the full framework including two stages depicted in (a) and (b). (a) an initial estimation of the wave field $\eta(\boldsymbol{x},t_0)$ using the RAO-based method, which is assembled into the initial state vector $\mathbf{q}(t_0)=\left[\eta(\boldsymbol{x},t_0),\psi(\boldsymbol{x},t_0),\boldsymbol{s}(t_0),\dot{\boldsymbol{s}}(t_0)\right]$. (b) reconstruction of $\eta(\boldsymbol{x},t)$ with $t>t_0$ by EnKF data assimilation, where an ensemble of simulations are carried over from $t_j$ to $t_{j+1}$ to generate $\mathbf{q}^{f}_{j+1}$, and assimilation of motion data for Kalman correction ($\mathbf{q}^{f}_j \rightarrow \mathbf{q}^{a}_j$) is conducted at each $t_j$ ($j\ge 1$). The mean state, as the solution of reconstruction, is denoted by (\fullredthick). The true solution is denoted by (\fullblackthick). }     
    
    \label{fig:schm method}
\end{figure}
\begin{algorithm}[h]
		\SetAlgoLined
		\textbf{Step 1}: RAO-based initial condition\;
		\Indp
        \textbf{Inputs}: $s^l_\text{m}(t)$ (body motion time history), $\rv{\theta(k)}^0$ and $\rv{a(k)}^0$ (initial guess of wave direction and amplitude), $\mathscr{K}$ (the set of wavenumber magnitudes)\;
        \textbf{Begin}\;
        \Indp Compute $\hat{s}^l_{\text{m},\omega}$ from motion history $s^l_\text{m}(t)$ using FFT\; 
         For $k \in \mathscr{K}$:\\
		\Indp Solve for  $(\rv{a(k)},\rv{\theta(k)})$ by minimizing (\ref{eq:RAO loss}); \\
        \Indm Compute  $\eta(\boldsymbol{x},t)$ using (\ref{eq: eta dominant dir});\\ 
        Evaluate $\eta(\boldsymbol{x},t_0)$.\\
        \Indm
        \textbf{End}\\
		\Indm
		\textbf{Step 2}: EnKF data assimilation\;
		\Indp
        \textbf{Input}: $\eta(\boldsymbol{x},t_0)$ (initial surface elevation), $N_T$ (number of assimilation steps), $N_e$ (number of ensemble members), $\boldsymbol{\theta}_0$ (initial guess of uncertain parameter), $\epsilon_p$ (threshold for loss (\ref{eq: expectation}));\\
        \textbf{Begin}\\
        \Indp
		Assemble $\overline{\mathbf{q}}_{j=0}=[\eta(\boldsymbol{x},t_0),\psi(\boldsymbol{x},t_0),\boldsymbol{s}({t_0}),\dot{\boldsymbol{s}}({t_0})]^{\top}$. \\
        Generate $\{\mathbf{q}^{n}_{j=0}\}$ for $n=\{1,...,N_e\}$; \\
        \textbf{For} $j\in\{1,2,...,N_T\}$:\\
        \Indp
        Assign $\boldsymbol{\theta}_{j-1}$ to $\boldsymbol{\theta}^{(\iota=0)}$;\\
        \textbf{While} $\tilde{L}(\boldsymbol{\theta}^{(\iota+1)}|\boldsymbol{\theta}^{(\iota)})>\epsilon_p$:\\
		\Indp Compute $\{\mathbf{q}^{n}_{j}\}$  from $\{\mathbf{q}^{n}_{j-1}\}$ using (\ref{eq: HOS}) and (\ref{eq: Cummins first order}) with $\boldsymbol{\theta}^{(\iota)}$;\\
        Assign $\{\mathbf{q}^{n}_{j}\}$ to $\{\mathbf{q}^{\text{f},n}_{j}\}$, and compute $\{\mathbf{q}^{\text{a},n}_{j}\}$ using (\ref{eq: analysis})\;
        Assign $\{\mathbf{q}^{\text{a},n}_{j}\}$ to $\{\mathbf{q}^{n}_{j}\}$;\\
        Obtain $\boldsymbol{\theta}^{(\iota+1)}$ by maximize (\ref{eq: expectation}); \\
        $\iota = \iota + 1$\\
        \Indm
        \textbf{End}\\
        Assign $\boldsymbol{\theta}^{(\iota)}$ to $\boldsymbol{\theta}_j$\\
        $j=j+1$\\
        \Indm
        \textbf{End}\\
		\Indm
        \textbf{End}\\
        \Indm
		\caption{Ship-as-a-buoy algorithm}
		\label{alg:overall}
\end{algorithm}

The ``ship as a buoy’’ problem we consider in this paper can be formally stated as follows: On a two-dimensional horizontal surface  $\boldsymbol{x}=(x,y)$, we have $N$ ships located on $\boldsymbol{x}_j=(x_j,y_j)$, $j=1,…,N$. 
\rv{These ships have zero forward speed and oscillate around $\boldsymbol{x}_j$ passively.} 
For each ship, measured time series of ship motion $\boldsymbol{s}^l_\text{m}(t)$ is available, where $l\in \mathcal{M}$ with $\mathcal{M}$ being a subset of the set $\{1,2,3,4,5,6\}$ giving the degrees of freedom which are measured.
Our goal is to reconstruct the ocean surface $\eta(\boldsymbol{x},t)$ in both space and time. We also consider parameters associated with ship dynamics, e.g., the added mass, RAOs, to be known, although deviations from their true values are allowed, which will be corrected as part of the solution.

In this section, we introduce the formulation and computational methods developed to solve the above problem. A schematic overview of the framework is illustrated in figure \ref{fig:schm method}, which we describe first followed by detailed descriptions of each component in each sub-section. The full algorithm is divided into two steps. In the first step with $t\in[0,t_0]$ (figure \ref{fig:schm method}(a)), we use the time series $\boldsymbol{s}(t)$ to reconstruct the wave field $\eta(\boldsymbol{x}, t)$ through our new RAO-based approach (described in detail in \S \ref{subsec:initial guess from RAO}$(a)$). The reconstructed $\eta(\boldsymbol{x}, t_0)$, as well as $\psi(\boldsymbol{x}, t_0)$ constructed using an approach in \citet{wang2021phase}, are used as an informed initial condition to start the second step. In the second step (figure \ref{fig:schm method}$(b)$), we treat the wave-body interaction as a coupled dynamical system with state vector $\mathbf{q}=[\boldsymbol{\eta},\boldsymbol{\psi},\boldsymbol{s},\dot{\boldsymbol{s}}]^{\top}$, where $\boldsymbol{\eta}$ and $\boldsymbol{\psi}$ represent surface wave elevation and velocity potential at all grid points, and $\boldsymbol{s}$ and $\dot{\boldsymbol{s}}$ are the six DoF displacement and velocity of ship motion, respectively.  The coupled system is evolved in time starting from $\mathbf{q}(t_0)$ using high-order spectral method for $\boldsymbol{\eta}$ and $\boldsymbol{\psi}$ (\S \ref{subsec:Ocean surface wave}), and Cummins equation for $\boldsymbol{s}$ and $\dot{\boldsymbol{s}}$ (\S \ref{subsec:floating body dynamics}). At $t=t_j$ when the measured ship motions are available, we apply an ensemble Kalman filter (EnKF) to obtain an optimal correction to $\mathbf{q}(t_j)$ (\S \ref{subsec:EnKF}), with the corrected result (i.e., analysis result) serving as initial condition for subsequent simulation. We note that for simplicity, figure \ref{fig:schm method} does not include estimation of uncertain parameters in the floating body dynamics, which will be presented separately in \S \ref{subsec: uncertain parameters}. A pseudo code of the full process is presented in the algorithm \ref{alg:overall}, which can be used to help understand the details presented in the following subsections.


\subsection{Initial guess from RAO}\label{subsec:initial guess from RAO}

We recall that RAOs for a floating body are complex functions defined as
\begin{equation}\label{eq: RAO}
     Z^l(\boldsymbol{k})=\frac{r^l_{\boldsymbol{k}}}{a_{\boldsymbol{k}}}
\end{equation}
where $r^l_{\boldsymbol{k}}$ with $l=\{1,2,...,6\}$ is the Fourier coefficient of the $l^{\text{th}}$ DoF motion induced by a \rv{wave mode $a_{\boldsymbol{k}}\mathrm{e}^{i\omega t - i\boldsymbol{k}\cdot\boldsymbol{x}}$ with complex amplitude $a_{\boldsymbol{k}}$
}. 
Specifically, $a_{\boldsymbol{k}}$ and $r^l_{\boldsymbol{k}}$ are defined such that the incident wave profile $\eta(\boldsymbol{x},t)$ and motion time series $s^l(t)$ can be expressed as
\begin{equation}
   \eta(\boldsymbol{x},t)=\sum_{\boldsymbol{k}}\Re\{a_{\boldsymbol{k}} \mathrm{e}^{i\omega t-i\boldsymbol{k}\cdot\boldsymbol{x}}\},  
\end{equation}
\begin{equation}\label{eq: measured signal representation}
    s^l(t)=\sum_{\boldsymbol{k}}\Re\{r^l_{\boldsymbol{k}}\mathrm{e}^{i\omega t}\} = \sum_{\boldsymbol{k}}\Re\{Z^l(\boldsymbol{k})a_{\boldsymbol{k}}\mathrm{e}^{i\omega t}\}, 
\end{equation}
with $\omega=\sqrt{gk}$ the \rv{deep-water} dispersion relation and $k=|\boldsymbol{k}|$ the scalar wavenumber. The goal is to reconstruct $\eta(\boldsymbol{x},t)$ from the measured time series $s^l_{\text{m}}(t)$ with $l \in \mathcal{M}$, associated with Fourier coefficient $\hat{s}^{l}_{\mathrm{m},\omega}$. For each fixed $\omega$, one need to invert a linear system (one can think of taking the frequency Fourier transform of (\ref{eq: measured signal representation})), 
\begin{equation}
\hat{s}^l_{\text{m},\omega}= \sum_{\boldsymbol{k}:\,\omega(k)\in \Omega} Z^l(\boldsymbol{k})a_{\boldsymbol{k}},
\end{equation}
where $\Omega$ represents a small interval around $\omega$ depending on the discretization in $\omega$ to generate $\hat{s}^{l}_{\mathrm{m},\omega}$ (say as the inverse of the length of $s^l_{\text{m}}(t)$). This is clearly an under-determined system as discussed in \S \ref{sec: introduction}, because for each $\omega$ there are a large number of $\boldsymbol{k}$ satisfying $\omega(k)\in \Omega$ (say due to waves traveling in many directions) but at most 6 (since $|\mathcal{M}|\leq 6$) equations to solve.

To overcome the under-determinedness, we introduce a critical but reasonable assumption: We assume that the response at each $\omega$ is induced by a single dominant wave component in direction \rv{$\theta(k)$}, which is to be determined. This simply means that the wave field is composed of components of different frequencies, with each frequency component represented by one wave traveling in one ``dominant’’ direction \rv{$\theta(k)$} with amplitude \rv{$a(k)$} (not to be connected to $a_{\boldsymbol{k}}$ as this is simply a new wave field definition).
The total wave field can therefore be expressed as
\begin{equation}\label{eq: eta dominant dir}
\eta(\boldsymbol{x},t)=\sum_k \Re\{ \rv{a(k)} \mathrm{e}^{i\omega t- i\boldsymbol{k}\cdot\boldsymbol{x}} \}, \mathrm{with}\,\, \boldsymbol{k}=\rv{(k\cos\left(\theta\left(k\right)\right), k\sin\left(\theta\left(k\right)\right))}.
\end{equation}

The problem can now be accordingly reformulated: for each $\omega$, we need to invert
\begin{equation}\label{eq: measured signal single wave}
    \hat{s}^l_{\mathrm{m},\omega} = Z^l(k,\rv{\theta(k)})\rv{a(k)}, 
\end{equation}
with two unknowns, $\rv{\theta(k)}$ real and $\rv{a(k)}$ complex, but $|\mathcal{M}|$ complex equations (up to 6). The problem is no longer under-determined (and may become over-determined).
We note that for the special case of $|\mathcal{M}|=1$, (\ref{eq: measured signal single wave}) remains under-determined and to some extent ill-posed. This is, however, consistent with the underlying ill-defined physical problem. For example, if only heave motion is known, there is no way to determine the direction of the incident wave. 

We invert (\ref{eq: measured signal single wave}) through a nonlinear least square approach, by minimizing the loss function defined as 
\begin{equation}\label{eq:RAO loss}
    \mathcal{J}_{\omega(k)}(a(k),\rv{\theta(k)}) = \frac{1}{2}\sum_{l\in \mathcal{M}}\left(\hat{s}^l_{\mathrm{m},\omega} - Z^l(k,\rv{\theta(k)})\rv{a(k)}\right)\left(\hat{s}^l_{\mathrm{m},\omega} - Z^l(k,\rv{\theta(k)})\rv{a(k)}\right)^{*},
\end{equation}
where $(\cdot)^{*}$ represents the complex conjugate. We adopt an L-BFGS optimizer to find the minimizer of (\ref{eq:RAO loss}) using the analytical gradient (see Appendix \ref{sec:RAO_loss_gradient} for details, where $dZ^l/d\rv{\theta(k)}$ is expressed as a function of hydrodynamic coefficients such as frequency dependent added mass $\boldsymbol{A}(\omega)$ and damping matrix $\boldsymbol{B}(\omega)$). Compared to the conventional RAO-based methods such as those in \citet{nielsen2006estimations,iseki2000bayesian}, our approach significantly reduces the dimensionality of the optimization problem (from many unknowns to just two: $\rv{\theta(k)}$ (real) and $\rv{a(k)}$ (complex), as seen from (\ref{eq: measured signal representation}) to (\ref{eq: measured signal single wave})). In the meanwhile, our problem becomes nonlinear due to the introduction of unknown $\rv{\theta(k)}$, but this does not cause significant computational cost due to the availability of analytical gradients. In addition, our approach does not require the regularization in \citet{nielsen2006estimations}, whose strength determination is likely the most expensive component in the previous approaches. 

Because of the above features, the approach we develop here is  orders-of-magnitude faster than those in \citet{nielsen2006estimations}, favoring their application in real time. In the meanwhile, it provides sufficient accuracy for our purpose as we will show in 
\S \ref{subsec: results RAO 1 body}. For simplicity, we have presented above the method based on one single ship, but it is straightforward to extend it to situations with multiple ships. One simply need to form the RAO-based equation in (\ref{eq: measured signal single wave}) for different ships and add the corresponding loss terms in (\ref{eq:RAO loss}) as we will also demonstrate in \ref{subsec: multiple ships}. 

Due to the assumption regarding dominant wave traveling direction, the algorithm works better for wave fields that are relatively long-crested. \rv{Therefore, we mainly consider long-crested waves in this paper, with one set of results for short-crested waves presented in the appendix \ref{sec: appendix large beta}.} In principle, to \rv{accurately} deal with short-crested wave field, one can add more unknown parameters by including multiple directional wave components for each wavenumber magnitude $k$, i.e., including more terms in (\ref{eq: eta dominant dir}). We leave this extension to future work.

\subsection{Surface wave prediction by HOS}\label{subsec:Ocean surface wave}
\del{\begin{itemize}
    \item Zakharov equation, HOS
\end{itemize}}

We consider the two-dimensional free surface of an incompressible and inviscid fluid flow, in the $x$-$y$ plane, with $z$ representing the vertical direction. The free surface motion is governed by the Euler equation in Zakharov form \citep{zakharov1968stability} representing the kinematic and dynamic boundary conditions of the fluid surface respectively, 
\begin{subequations}\label{eq: HOS}
    \begin{align}
 \frac{\partial \eta}{\partial t}+\frac{\partial \psi}{\partial \boldsymbol{x}} \cdot \frac{\partial \eta}{\partial \boldsymbol{x}}-\left[1+\frac{\partial \eta}{\partial \boldsymbol{x}} \cdot \frac{\partial \eta}{\partial \boldsymbol{x}}\right] \phi_z&=0, \\
\frac{\partial \psi}{\partial t}+\frac{1}{2} \frac{\partial \psi}{\partial \boldsymbol{x}} \cdot \frac{\partial \psi}{\partial \boldsymbol{x}}+g\eta-\frac{1}{2}\left[1+\frac{\partial \eta}{\partial \boldsymbol{x}} \cdot \frac{\partial \eta}{\partial \boldsymbol{x}}\right] \phi_z^2&=0, 
\end{align}
\end{subequations}
The surface elevation $\eta$ and surface potential $\psi$ are both functions of horizontal coordinate $\boldsymbol{x}$ and $t$, and the latter is related to the bulk velocity potential $\phi(\boldsymbol{x},z,t)$ by $\psi(\boldsymbol{x},t)=\phi(\boldsymbol{x},\eta,t)$. The vertical velocity $\phi_z = \partial \phi/\partial z|_{z=\eta}$ is evaluated at the surface location, and is related to the surface potential $\psi$ through a Dirichlet-to-Neumann problem with a field satisfying a Laplace equation. 
In HOS, we solve the Dirichlet-to-Neumann problem using an efficient pseudo-spectral method \citep{dommermuth1987high} and integrate the equation in time with $4^{\text{th}}$-order Runga-Kutta method.

\subsection{Body motion prediction by Cummins equation}\label{subsec:floating body dynamics}
\del{
\begin{itemize}
    \item Time domain Cummins equation, Froude Krylov force. Mention the limitation of Cummins equation: long wave
    \item Cummins equation is only an demonstration of the body dynamics. Other models also work. 
    \item (Move to RAO) Frequency space formulation, Response amplitude operator
    \item A schematic figure showing the floating body on a wave field, not to violate those assumptions. 
\end{itemize}
}

We consider a floating body oscillating around its equilibrium position under wave excitations, as illustrated in figure \ref{fig:wave body schm}$(a)$. The instantaneous displacement of the body is described by a vector $\boldsymbol{s}$ with six components, including surge ($x$-translation), sway ($y$-translation), heave ($z$-translation), roll (rotation around $x$-axis), pitch (rotation around $y$-axis) and yaw (rotation around $z$-axis). One commonly used model for the dynamics of the floating body is the Cummins equation \citep{Cummins1962,ogilvie1964recent}, 
\begin{equation}\label{eq:Cummins}
    \begin{aligned}
\left(\boldsymbol{M}+\boldsymbol{M}_a\right) \frac{\mathrm{d}^2\boldsymbol{s}}{\mathrm{d}t^2}(t) +\int_0^t \boldsymbol{K}(t-\tau) \frac{\mathrm{d}\boldsymbol{s}}{\mathrm{d}t}(\tau) \mathrm{d} \tau +\boldsymbol{C}^R \boldsymbol{s}(t)=\boldsymbol{f}(t), 
\end{aligned}
\end{equation}
where $\boldsymbol{M},\boldsymbol{M}_a \in \mathbb{R}^{6\times 6}$ represent the mass and added mass matrices of the body. The time-dependent matrix $\boldsymbol{K}(t)$ corresponds to the fluid memory effect caused by radiation-induced hydrodynamic forces \citep{fossen2011handbook}. 
$\boldsymbol{C}^R$ is the restoring matrix, with the corresponding term accounting for the restoring force generated from the buoyancy effect. Details about these matrices can be found in Appendix \ref{sec: appendix Cummins}. The right-hand-side term $\boldsymbol{f}(t)$ is the wave excitation force, which is dominated by the Froude-Krylov force under long wave approximation (i.e., size of the body much smaller than the typical wavelength):
\begin{equation}\label{eq: Froude Krylov force}
    \boldsymbol{f}=-\int_{A_w} \left(\begin{array}{c}
p\boldsymbol{n} \\
p\boldsymbol{r}\times\boldsymbol{n}
\end{array}\right)\mathrm{d}A,
\end{equation}
where $\boldsymbol{n}$ is the unit surface normal vector pointing into the fluids,  and $\boldsymbol{r}$ is the radial vector pointing from the body gravity center to the body surface. The  integral in (\ref{eq: Froude Krylov force}) is over the wet surface of the body $A_w$. The wave induced pressure $p$ is computed on the nominal body surface, 
\begin{equation}
p(\boldsymbol{x},z,t)=\sum_{\boldsymbol{k}}\rho g\, \tilde{\eta}_{\boldsymbol{k}}(t) \,e^{kz+i\boldsymbol{k}\cdot\boldsymbol{x}}\,
\end{equation}
where $\tilde{\eta}_{\boldsymbol{k}}(t)$ is the Fourier coefficient of $\eta(\boldsymbol{x},t)$ obtained from the results of HOS simulations. 
To better fit the state-space model that is widely applied in ensemble Kalman filter, we transform (\ref{eq:Cummins}) into a first order system, 
\begin{subequations}\label{eq: Cummins first order}
    \begin{align}
\frac{\mathrm{d}\boldsymbol{s}}{\mathrm{d}t}(t)&=\boldsymbol{v}, \\
\left(\boldsymbol{M}+\boldsymbol{M}_a\right) \frac{\mathrm{d}\boldsymbol{v}}{\mathrm{d}t}(t) &=-\int_0^t \boldsymbol{K}(t-\tau) \boldsymbol{v}(\tau) \mathrm{d} \tau -\boldsymbol{C}^R \boldsymbol{s}(t)+\boldsymbol{f}(t),  
\end{align}
\end{subequations}
where $\boldsymbol{v}$ is the six-DoF body velocity. The system (\ref{eq: Cummins first order}) is solved using a $4^{\text{th}}$-order Runge-Kutta scheme in time in the present work. 

We note that in addition to the long wave approximation, the present implementation of Cummins equation also assumes small body motion. It is certainly possible to relax these assumptions, say, by adding a diffraction force to $\boldsymbol{f}(t)$, or even replacing the Cummins equation by more sophisticated models. Due to the non-intrusive nature of the ensemble-based framework, these potential ``improvement’’ can be included in a natural way without affecting other components of the framework.

Finally, it is worthwhile to point out that the Cummins equation (\ref{eq:Cummins}) and the RAO provide the same information for a floating body --- they can be considered as Fourier pairs \citep{fossen2011handbook}. To see this, we can take a frequency transformation of (\ref{eq:Cummins}), leading to
\begin{equation}
    \frac{\boldsymbol{r}_{\boldsymbol{k}}}{a_{\boldsymbol{k}}}=
    \left(-\omega^2(\boldsymbol{M}+\boldsymbol{A}(\omega))+i\omega \boldsymbol{B}({\omega})+\boldsymbol{C}^R\right)^{-1}\int_{A_w} \left(\begin{array}{c}
\boldsymbol{n} \\
\boldsymbol{r}\times\boldsymbol{n}
\end{array}\right)\rho g\,  \,e^{kz-i\boldsymbol{k}\cdot\boldsymbol{x}}\,\mathrm{d}A(\boldsymbol{x},z),
\end{equation}
which gives an expression of (\ref{eq: RAO}) via hydrodynamic coefficients of the floating body. The frequency-dependent added mass $\boldsymbol{A}(\omega)$ and the damping coefficient $\boldsymbol{B}(\omega)$ are related to the matrices in the original Cummins equation (\ref{eq:Cummins}) through, 
\begin{equation}
    \boldsymbol{A}(\omega) = \boldsymbol{M}_a - \frac{1}{\omega}\int_0^{\infty}\hat{\boldsymbol{K}}(\omega)\sin(\omega\tau)\mathrm{d}\tau, \qquad \boldsymbol{B}(\omega) = \int_0^{\infty}\hat{\boldsymbol{K}}(\omega)\cos(\omega\tau)\mathrm{d}\tau, 
\end{equation}
where $\hat{\boldsymbol{K}}(\omega)$ is the element-wise Fourier transform of $\boldsymbol{K}(t)$.

\subsection{Ensemble Kalman filter}\label{subsec:EnKF}
\del{
\begin{itemize}
    \item The data assimilation setup, state vector, observation operator
    \item The update equations for EnKF, 
\end{itemize}
}
We recall the definition of the state vector $\mathbf{q}$ as 
\begin{equation}
    \mathbf{q}= \left[\begin{array}{c}
\boldsymbol{\eta} \in \mathbb{R}^N \\
\boldsymbol{\psi} \in \mathbb{R}^N\\
\boldsymbol{s} \in \mathbb{R}^6 \\
\boldsymbol{\dot{s}} \in \mathbb{R}^6
\end{array}\right]. 
\end{equation}
At given time $t$, $\mathbf{q}(t)$ includes the wave elevation and velocity potential at $N=N_x\times N_y$ grid points, as well as six displacements and velocities of the body. 
We then consider an ensemble of perturbed states $\{\mathbf{q}^n\}$, $n=1,2,...,N_e$, with each member perturbed from the mean state $\overline{\mathbf{q}} = \sum_{n=1}^{N_e}\mathbf{q}^n/N_e$. From these definitions, we also construct a perturbation matrix
\begin{equation}
    \mathbf{Q} = [\mathbf{q}^1-\overline{\mathbf{q}},\mathbf{q}^2-\overline{\mathbf{q}},...,\mathbf{q}^{N_e}-\overline{\mathbf{q}}]
\end{equation}
At $t_0$, the wave components of $\overline{\mathbf{q}}$ are initialized by the linear wave solution constructed in \S \ref{subsec:initial guess from RAO}; the body components of $\overline{\mathbf{q}}$ are initialized by the measured instant body displacement and velocity for the measured DoFs, and set up to zero otherwise. The wave components of $\mathbf{q}^n$ are then constructed by adding random perturbations to $\overline{\mathbf{q}}$ following the approach described in \citet{wang2021phase}. The body displacement and velocity components of  $\mathbf{q}^n$ are constructed by adding independent Gaussian noise to each DoF, i.e., $\boldsymbol{s}^{n} = \overline{\boldsymbol{s}} + \boldsymbol{\epsilon}^n$ and $\dot{\boldsymbol{s}}^{n} = \overline{\dot{\boldsymbol{s}}} + \boldsymbol{\epsilon}^n$, where $\boldsymbol{\epsilon}^n\sim N(0,\mathbf{R})$ with $\mathbf{R}$ a $6\times6$ diagonal matrix with numerical values further provided in \S \ref{sec:results}. We note that in principle one can set up different Gaussian noises for $\boldsymbol{s}^{n}$ and $\dot{\boldsymbol{s}}^{n}$, but the tests conducted in \S \ref{sec:results} show that the current setup already yields satisfactory results.   


Subsequently, we apply EnKF for a number of iterations until desired analysis result is achieved. Each iteration contains a forecast step from $t_{j-1}$ to $t_j$, and an analysis step at $t_j$. In the forecast step, we compute $\mathbf{q}_j^n=\mathcal{N}(\mathbf{q}_{j-1}^{n}, \dot{\boldsymbol{s}}_P)$ where $\mathcal{N}$ represents the time marching of the wave-body coupled system through HOS and Cummins equation, and $\dot{\boldsymbol{s}}_P$ represents $\dot{\boldsymbol{s}}$ at a sufficient number of previous time steps to compute the integral term in the Cummins equation (\ref{eq: Cummins first order}). Although this system specified by $\mathcal{N}$ is not a standard state-space model that can only depend on $\mathbf{q}_{j-1}^{n}$, the relaxation of this assumption does not pose difficulty in the assimilation. 
We then assign the result of the forecast step $\mathbf{q}_j^n$ into $\mathbf{q}^{f,n}_j$ and assemble the corresponding matrix $\mathbf{Q}^{f}_j$. In the analysis step at $t_j$, we take the measurement $\boldsymbol{s}_{\text{m},j}$, and {perturb it by $\boldsymbol{s}^{n}_{\mathrm{m},j} = \boldsymbol{s}_{\mathrm{m},j} + \boldsymbol{\epsilon}_{\text{m},j}^n$ to form $N_e$ ensemble members, where 
$\boldsymbol{\epsilon}_{\text{m},j}^n\sim N(0,\mathbf{R}_\text{m})$ with $\mathbf{R}_\text{m}$ a $|\mathcal{M}|\times|\mathcal{M}|$ sub-matrix of $\mathbf{R}$ containing only the measured DoFs. The ensemble Kalman filter can then be applied through the following analysis equation \citep{evensen2003ensemble}, 
\begin{equation}\label{eq: analysis}
    \mathbf{q}^{\mathrm{a},n}_j=\mathbf{q}^{\mathrm{f},n}_j+\mathbf{K}_j\left[\boldsymbol{s}^{n}_{\mathrm{m},j}-\mathbf{H} \mathbf{q}^{\mathrm{f},n}_j\right], 
\end{equation}
where $\mathbf{H}$ is the observation operator that maps the full state vector to the observed body displacements. The Kalman gain $\mathbf{K}_j$ is defined as,  
\begin{equation}
    \mathbf{K}_j=\left(\mathbf{Q}^{\mathrm{f}}_j\right)\left(\mathbf{Q}^{\mathrm{f}}_j\right)^{\top} \mathbf{H}^{\top}\left[\mathbf{H}\left(\mathbf{Q}^{\mathrm{f}}_j\right)\left(\mathbf{Q}^{\mathrm{f}}_j\right)^{\top} \mathbf{H}^{\top}+\mathbf{R}_\text{m}\right]^{-1}, 
\end{equation}
At the end of the analysis step, we assign $\mathbf{q}^{\text{a},n}_j$ into $\mathbf{q}^{n}_j$ and continue with the next forecast step for iterations.

We remark that the Kalman gain maps the observational misfit to the optimal correction of the full state vector. From a Bayesian inference perspective, the analysis step (\ref{eq: analysis}) can be interpreted as combining the prior estimation $\mathbf{q}^{\mathrm{f},n}_j$ with the data likelihood (Kalman gain increment) to provide the posterior corrected state $\mathbf{q}^{\mathrm{a},n}_j$. For a more detailed and systematic review of ensemble Kalman filter, see \citet{evensen2003ensemble, katzfuss2016understanding}.

\subsection{Maximum likelihood estimation of uncertain parameters }\label{subsec: uncertain parameters}
The parameters in floating body models, such as those in the Cummins equation (\ref{eq:Cummins}), may contain errors in their pre-computed values, when compared to real values of a ship at sea. Therefore, it is desirable for their values to be corrected online in the assimilation process. In general, there are two classes of methods available for parameter estimations in the EnKF process \citep{katzfuss2016understanding}: augmented state approach and maximum likelihood approach. \citet{wang2022phase} adopted the first one to estimate the unknown current parameters in the ocean wave model. In this work we adopt the second approach, which is not necessarily better, but does avoid the assumption of joint Gaussian distribution of the state and the parameters. 

To describe the maximum likelihood approach, we consider a general situation where a vector of uncertain parameters $\boldsymbol{\theta}$ exists in the dynamic model $\mathcal{N}_{\boldsymbol{\theta}}$, which needs to be corrected in each assimilation step. At the $j$-th step of assimilation, the optimal vector $\boldsymbol{\theta}^{*}$ corresponds to that which maximizes the likelihood of observation $\boldsymbol{s}_{\text{m},j}$, i.e., 
\begin{equation} \label{eq: optimal theta}
    \boldsymbol{\theta}^{*} = \mathrm{arg}\max_{\boldsymbol{\theta}} L(\boldsymbol{\theta}),\qquad L(\boldsymbol{\theta}) = \log p(\boldsymbol{s}_{\text{m},j} \mid \boldsymbol{\theta}), 
\end{equation}

To compute $ p(\boldsymbol{s}_{\text{m},j}\mid \boldsymbol{\theta})$, we notice
\begin{equation}
\label{eq: evidence parameter}
     p(\boldsymbol{s}_{\text{m},j} \mid \boldsymbol{\theta})= \int p(\boldsymbol{s}_{\text{m},j}, \mathbf{q}_{j} \mid \boldsymbol{\theta}) \mathrm{d}  \mathbf{q}_{j}
\end{equation}
where the state vector $\mathbf{q}_{j}$ is random (say from the ensemble representation) and is treated as a hidden variable. The integral over $\mathbf{q}_{j}$ can be considered as a marginalization over all possible states. 

A direct computation of $ p(\boldsymbol{s}_{\text{m},j}\mid \boldsymbol{\theta})$ by (\ref{eq: evidence parameter}) is expensive since it involves a high-dimensional integration. The situation is made more difficult when we consider the optimization problem (\ref{eq: optimal theta}) because $\boldsymbol{\theta}$ and $\mathbf{q}_{j}$ are intertwined, i.e., for each updated $\boldsymbol{\theta}$, the high-dimensional integration needs to be performed again. This high computational cost of the problem can be alleviated through the expectation-maximization (EM) method \citep{dempster1977maximum, mons2024bayesian}. At each assimilation step $j$, the algorithm starts from an initial guess $\boldsymbol{\theta}^{(0)}$, which comes from the optimal $\boldsymbol{\theta}$ at assimilation step $j-1$. The EM method then update $\boldsymbol{\theta}$ through iterations defined as (with $\iota$ starting from 0)
\begin{equation}\label{eq: maximization}
    \boldsymbol{\theta}^{(\iota+1)}=\text{argmax}_{\boldsymbol{\theta}} \tilde{L}(\boldsymbol{\theta}\mid\boldsymbol{\theta}^{(\iota)}),
\end{equation}
with 
\begin{equation}\label{eq: expectation}
\tilde{L}(\boldsymbol{\theta}\mid\boldsymbol{\theta}^{(\iota)})=\mathbb{E}_{\mathbf{q}_{j}\mid \boldsymbol{s}_{\text{m},j},\boldsymbol{\theta}^{(\iota)}}[\log p(\boldsymbol{s}_{\text{m},j}, \mathbf{q}_{j}\mid \boldsymbol{\theta}) ].
\end{equation}

The expression $\tilde{L}(\boldsymbol{\theta}\mid\boldsymbol{\theta}^{(\iota)})$ can be shown as a lower bound of the original loss $L(\boldsymbol{\theta})$ \citep{bishop2006pattern}, and the iteration in (\ref{eq: maximization}-\ref{eq: expectation}) is guaranteed to increase the likelihood, and under practical conditions, converge to the solution of the original problem (\ref{eq: optimal theta}) \citep{wu1983convergence}. Moreover, (\ref{eq: expectation}) can be computed by a Monte-Carlo simulation (instead of high-dimensional integration), where the sampling of $\mathbf{q}_{j}\mid \boldsymbol{s}_{\text{m},j},\boldsymbol{\theta}^{(\iota)}$ is equivalent to sampling the analysis ensemble $\{\mathbf{q}^{\text{a},n}\}$ after performing the forecast $\mathcal{N}_{\theta}$ and assimilation of data $\boldsymbol{s}_{\text{m},j}$. The detailed computation of (\ref{eq: expectation}) is presented in Appendix (\ref{sec: appendix EM}), including the calculation of $p(\boldsymbol{s}_{\text{m},j}, \mathbf{q}_{j} \mid \boldsymbol{\theta})$ and the analytical derivative $\partial \tilde{L}(\boldsymbol{\theta}\mid\boldsymbol{\theta}^{(\iota)})/\partial \boldsymbol{\theta}$ needed for a gradient-based optimization. The converged solution $\boldsymbol{\theta}$ is then assigned as the initial guess $\boldsymbol{\theta}^{(0)}$ at assimilation step $j+1$ and the sequential EnKF assimilation continues.

\section{Numerical results}\label{sec:results}
We present the numerical results for three cases: one floating body (\S \ref{subsec: results RAO 1 body}), multiple floating bodies (\S \ref{subsec: multiple ships}) and one floating body with uncertain hydrodynamic coefficients (\S \ref{subsec: results uncertain parameters}). The true wave field is take from a \rv{reference HOS simulation, starting from an initial condition sampled from} a JONSWAP spectrum of significant wave height $H_s=0.57m$ \rv{(unless otherwise specified)}, peak wave period $T_p=10.03s$ and enhancement factor $\gamma=3$, with a spreading function, 
\begin{equation}
    D(\theta)= \begin{cases}\frac{2}{\beta} \cos ^2\left(\frac{\pi}{\beta} \theta\right), & \text { for }-\frac{\beta}{2}<\theta<\frac{\beta}{2} \\ 0, & \text { otherwise }\end{cases}, 
\end{equation}
where the spreading angle $\beta=\pi/6$ \rv{(a case with larger $\beta$ is shown in appendix \S \ref{sec: appendix large beta})}. In the HOS computation, the physical wave field is mapped to a computational domain of $2\pi\times 2\pi$ discretized by $N_x\times N_y=64\times 64$, with the peak wavenumber located at $k_p=4$. 

\rv{The measurement data is generated by a reference simulation of the floating body motion driven by the excitation force from the true wave}. The floating body used in all cases is an oblate ellipsoid with semi-axes $a_x = a_y = 5m, a_z=2.5m$ (in the computational domain scaled as $a_x = a_y=0.05, a_z=0.025$) and density half of the water density. The mass matrices $\boldsymbol{M}$, $\boldsymbol{M}_a$ and the memory kernel $\boldsymbol{K}(t)$ are precomputed in a project of Wave Energy Converter SIMulator (WEC-SIM) \citep{Lawson2016WEC}, with the dataset available online 
\footnote{\url{https://wec-sim.github.io/WEC-Sim/}}. We note that the parameters $k_p a_x=0.20$ and $H_s/a_z=0.23$ are both small, indicating that the approximations of long wave and small body motion are both valid, as two assumptions in the Cummins equation model (\ref{eq:Cummins}) discussed in \S \ref{subsec:floating body dynamics}. Some sample time series of the 6DoF body motion computed by (\ref{eq:Cummins}) in the above wave field are presented in Appendix \ref{sec: appendix Cummins}. Unless otherwise specified, we assume that all six DoFs are measured, i.e., $\mathcal{M}=\{1,2,3,4,5,6\}$. The RAO-based initial condition is estimated using time series in $[0,t_0]$ with $t_0=156T_p$, \rv{although comparable accuracy can be achieved with a substantially smaller $t_0$.} The EnKF data assimilation is applied for $t>t_0$ with an ensemble size of $N_e=288$ for all cases.

We also describe here some details of perturbations to generate the ensembles, used for all cases in the paper. At $t_0$, the ensemble of $\eta(\boldsymbol{x},t_0)$ is generated by perturbing the RAO-based initial condition with a Gaussian process of correlation lengthscale $1/k_p$ and standard deviation $\sigma=0.0017H_s$. \rv{This choice follows \citet{wang2021phase} and is motivated by earlier works \citep{LonguetHiggins1953,merigaud2018incorporating} that model random waves using Gaussian processes with certain lengthscales.} The perturbations associated with body motions are generated by setting the diagonal elements of $\mathbf{R}$ with standard deviations $(10^{-7},10^{-7},10^{-6},10^{-7},10^{-6},10^{-8})$ in the computational scale. These values correspond to about $0.1 \%$ to $1\%$ of the corresponding body-motion displacement (and velocity) standard deviations. \rv{Note that these standard deviations can affect the convergence rate of the filter. A high noise level typically lead to the contamination of measurement information and slow convergence of the filter. 
On contrary, a low noise level results in over-confidence to the measurements, consequently the filter overlooks the prediction step (governed by the physical equations) and the assimilated state diverge from the true state. 
We find that $0.1\%$ to $1\%$ of the body motion is an effective range of the noise standard deviation in order to demonstrate the capability of our method.}

\rv{The two steps of the framework require different computational effort. The RAO-based initialization is light-weighted and can be finished in less than one minute on a serial personal computer. The computational cost of the whole framework is dominated by the EnKF data assimilation step, where the intensive computations including wave-body coupled simulations and the matrix operations in EnKF updates are carried out. }

\subsection{One floating body}\label{subsec: results RAO 1 body}
\del{
\begin{itemize}
    \item Set up of RAO computation, time window, 
    \item Loss reduction with respect to wave frequency
    \item Reconstructed time sequence at the ship location
    \item A single snapshot of wave height at a time step, RAO vs true. 
\end{itemize}
}
\begin{figure}
    \centering  
    \includegraphics[width=0.6\textwidth]{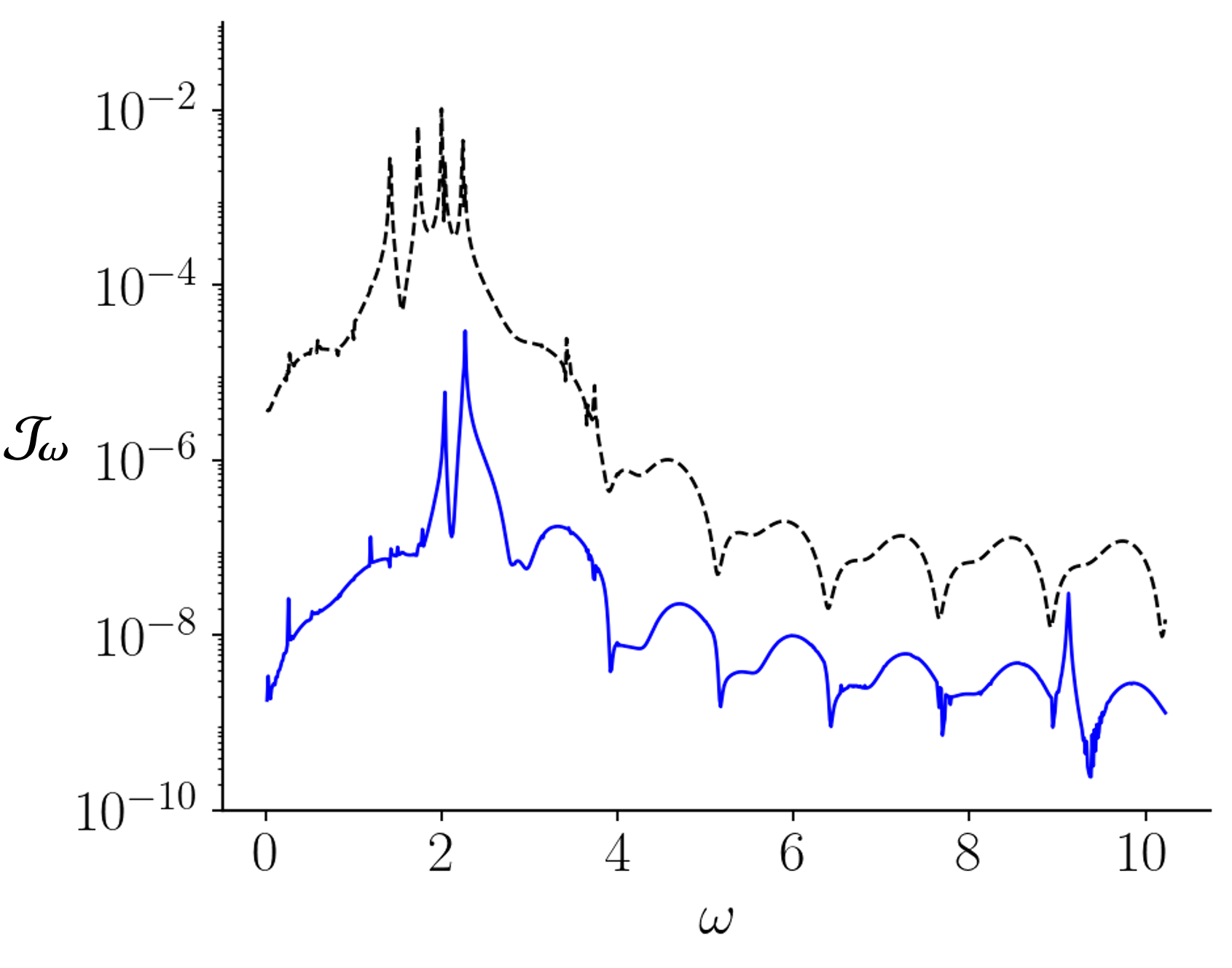}
    \caption{The loss $\mathcal{J}_\omega$ as a function of $\omega$ for initial guess of quiescent field (\dashedblack) and converged optimization result (\full).}
    \label{fig:RAO_loss}
\end{figure}

\begin{figure}
    \centering  \includegraphics[width=0.9\textwidth]{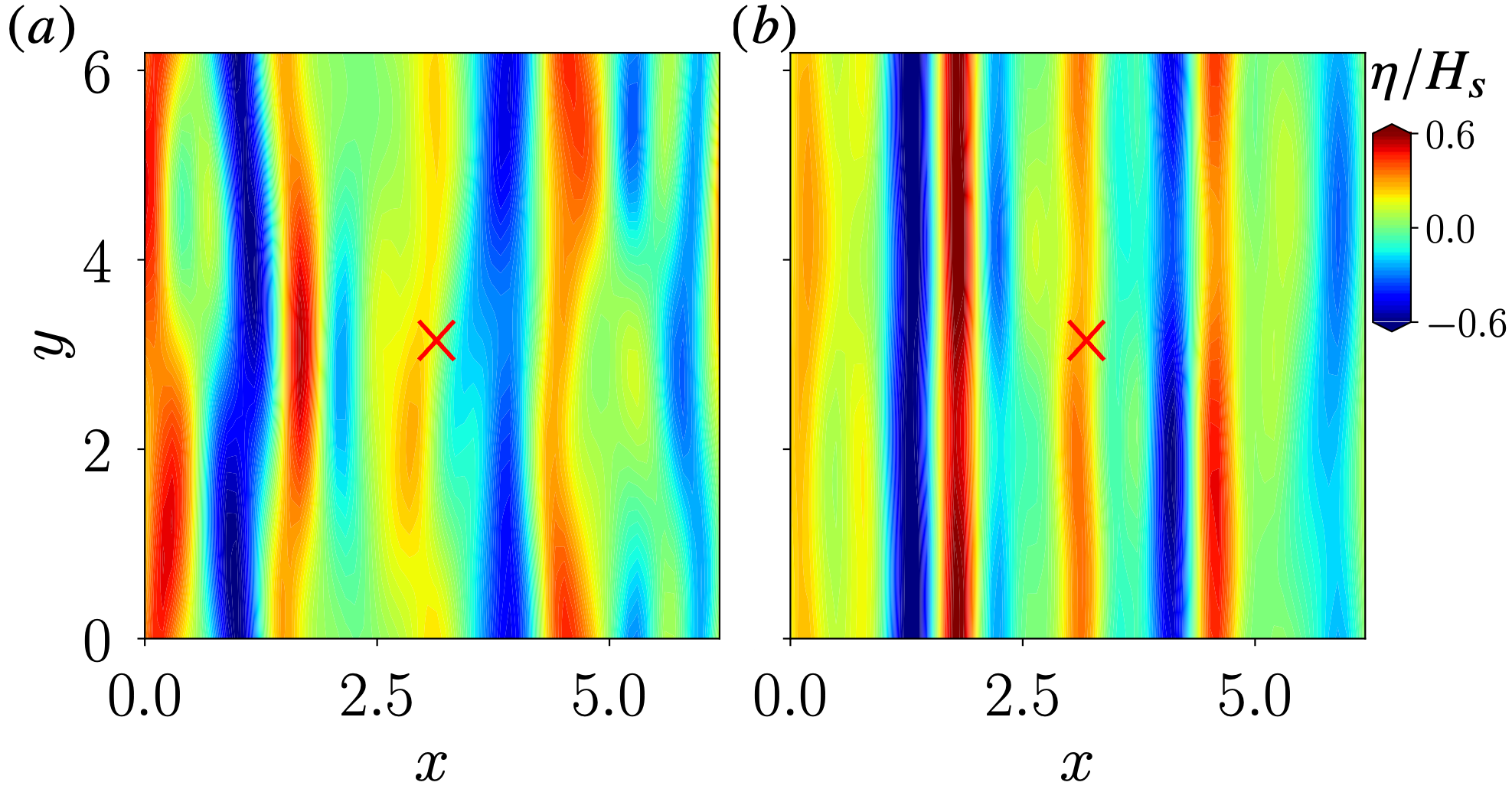}
    \caption{Comparison of $\eta(\boldsymbol{x},t_0)$ from (a) true solution and (b) RAO-based reconstruction, with the body location marked (\textcolor{red}{$\times$}). }
    \label{fig:RAO_vs_true_init}
\end{figure}

We consider one floating body located at $(x_b,y_b)=(\pi,\pi)$, the center of the domain. We first reconstruct a linear wave solution (to be used as initial condition later) from the time series $\boldsymbol{s}_\text{m}^l(t)$ with $l\in \mathcal{M}$ and $0<t<t_0\equiv 156T_p$, using the RAO-based approach described in \S \ref{subsec:initial guess from RAO}. In particular, we solve the optimization problem (\ref{eq:RAO loss}) starting from an initial guess of quiescent wave field, i.e., $\rv{a(k)}=0$ for all $k$. Figure \ref{fig:RAO_loss} plots the value of loss function (\ref{eq:RAO loss}) at different frequencies for the initial guess and converged solution, where the latter is reduced by 1-2 orders of magnitude indicating the effectiveness of our optimization algorithm. Figure \ref{fig:RAO_vs_true_init} shows the reconstructed wave field at $t=t_0$, compared to the true solution. It can be seen that the agreement is reasonable, but with some clear discrepancy. In general, the RAO-based solution is more ``long-crested'', which is due to the assumption of one ``dominant'' wave direction in our algorithm versus the true wave field with a finite spreading angle.

\begin{figure}
    \centering  \includegraphics[width=0.7\textwidth]{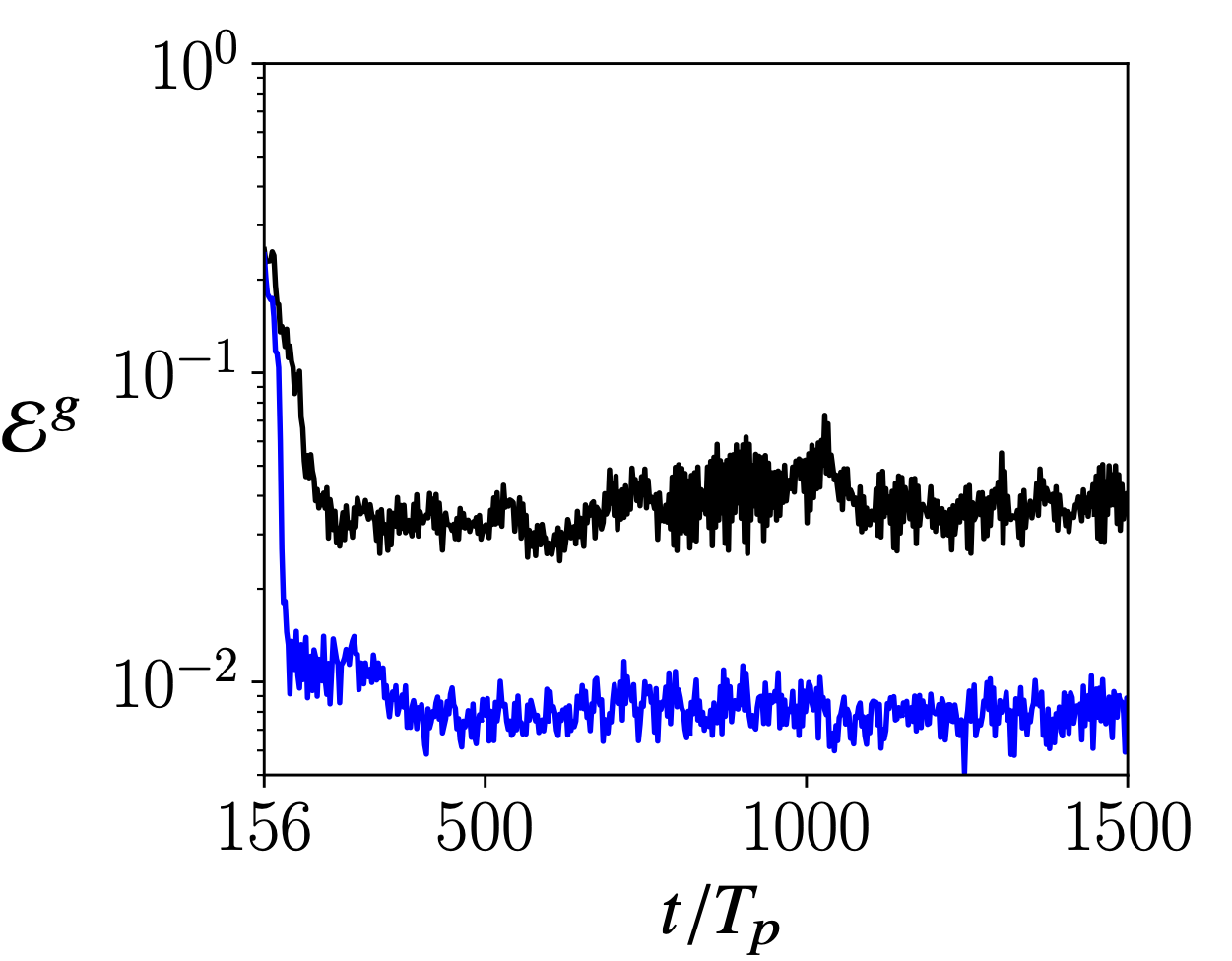}
    \caption{The global error $\mathcal{E}^g$ as a function of time, starting from zero initial condition (\fullblack{}) and RAO-based initial condition (\full). }
    \label{fig:RAO_vs_zero_error_history_compare}
\end{figure}

The RAO-based solution in figure \ref{fig:RAO_vs_true_init} is then used as an initial condition of the EnKF data assimilation process for $t>t_0$. To evaluate the performance of the EnKF reconstruction, we first define a normalized global error of the reconstructed wave field $\eta(\boldsymbol{x},t)$ as 
\begin{equation}
\mathcal{E}^g(t) =\frac{ \left(\overline{\left(\eta(\boldsymbol{x},t)-\eta^{\text{true}}(\boldsymbol{x},t)\right)^2}^s\right)^{1/2}}{H_s}, 
\end{equation}
where $\overline{(\cdot)}^s$ denotes a spatial average, and solution $\eta$ is taken as the ensemble average $\overline{\eta}$ in the EnKF framework. Figure \ref{fig:RAO_vs_zero_error_history_compare} plots $\mathcal{E}^g$ as a function of $t$, showing a significant reduction of the reconstruction error as EnKF proceeds, with a final error of $0.7\%$. In contrast, in figure \ref{fig:RAO_vs_zero_error_history_compare} we also include the error $\mathcal{E}^g(t)$ resulting from a zero initial condition (instead of the RAO-based solution). Not only do we see a higher initial error at $t=t_0$, but also the final stagnation error is higher, on the level of $4\%$. This comparison demonstrates the advantage to start the EnKF process from a favorable initial condition, i.e., the RAO-based solution in our case. This point is more clearly reflected in figure \ref{fig:snap_RAO_vs_zero}, where we show a comparison of the reconstructed wave fields at $t=t_0$ and final time $t=1700T_p$ to the true solution, for both the RAO-based and zero initial conditions. We find that at the final time $t=1700T_p$, the solution obtained by starting from the RAO-based initial condition is apparently closer to the true solution of the wave field.
\begin{figure}
    \centering  \includegraphics[width=1.0\textwidth]{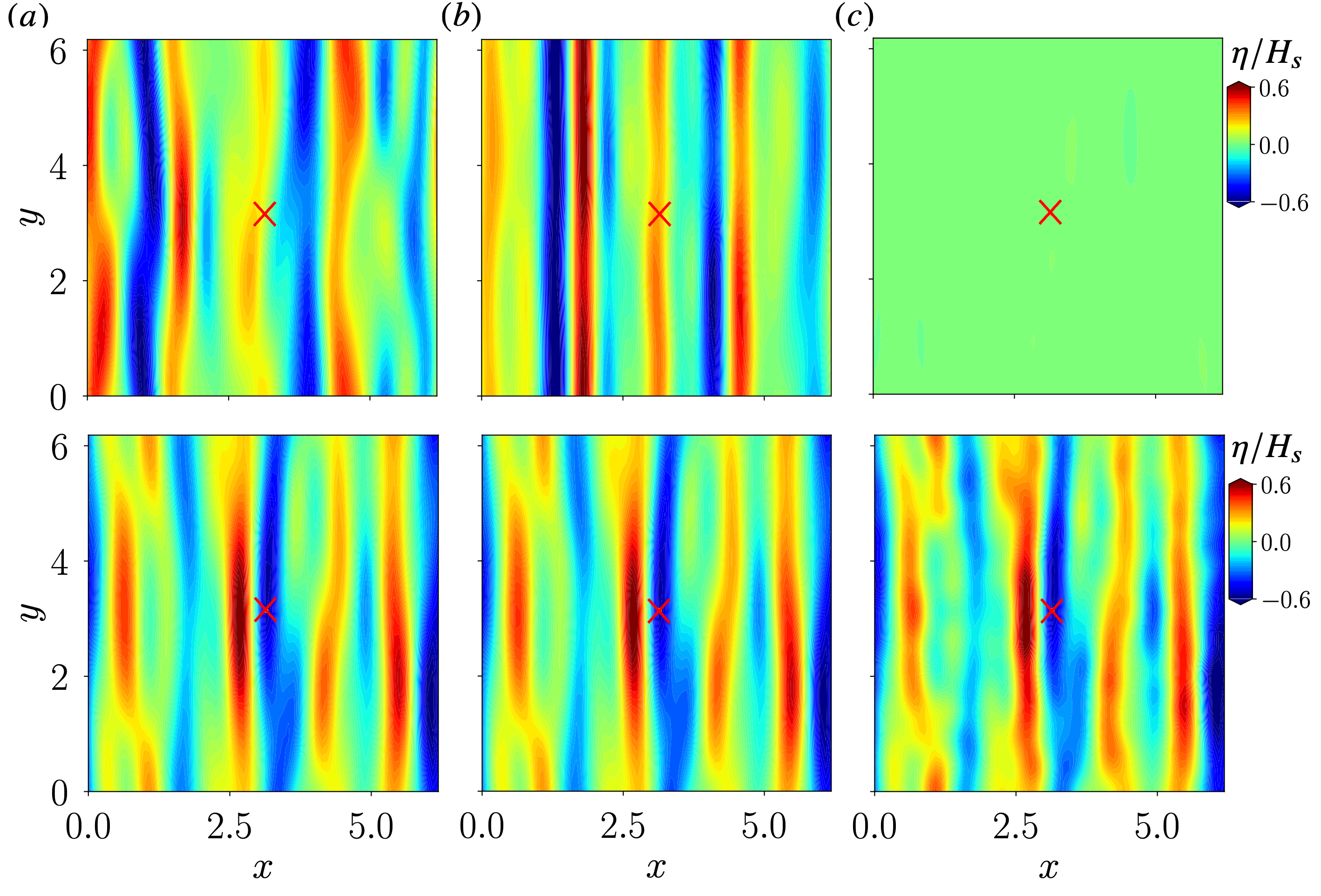}
    \caption{Comparison of $\eta$ at $t_0$ (top row) and final assimilation time (bottom row) among $(a)$ the true solution,  $(b)$ EnKF solution using RAO-based initial condition and $(c)$ EnKF solution using zero initial condition. The body location is marked (\textcolor{red}{$\times$}). }
    \label{fig:snap_RAO_vs_zero}
\end{figure}

\begin{figure}
    \centering  \includegraphics[width=0.7\textwidth]{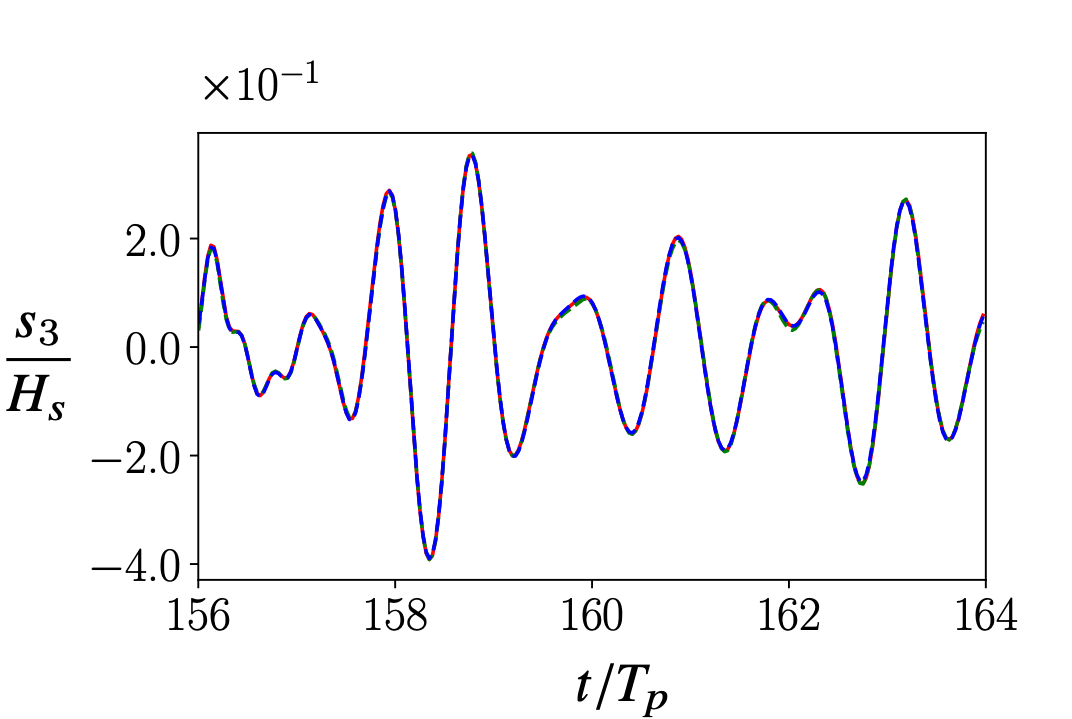}
    \caption{The time series $s_3(t)$ from true solution (\full), EnKF reconstruction with RAO-based initial condition (\dashedred) and with zero initial condition (\dashedgreen).}
    \label{fig:Z1Z2_RAO_0init_compare}
\end{figure}

We note that, although starting from a zero initial condition yields a less accurate wave-field reconstruction, the reconstructed motion time series remain satisfactory.
This is shown in figure \ref{fig:Z1Z2_RAO_0init_compare}, which plots the reconstructed motion time series ($l=3$ for example) starting from the RAO-based and zero initial conditions. 
Both cases show similar favorable agreement to the true solution for large $t$. 
This highlights the ill-posedness of the data assimilation problem: even when the measured degrees of freedom are reconstructed well (i.e., the measurements effectively correct the computed states), the inferred wave field need not be unique, and its accuracy may depend on the initial guess.
This phenomenon is similar to that observed in data assimilation for wall-bounded turbulence \citep{wang2022observable,wang2022synchronization}, where the reconstruction of interior flow from wall stress measurements might fail, even if the measured stress itself is reproduced correctly. Hereafter we only show results from the RAO-based initial condition.

To conduct a more thorough analysis on the reconstructed field, we further define a spatial error as
\begin{equation}\label{eq: error space}
\mathcal{E}(\boldsymbol{x}) =\frac{ \left(\overline{\left(\eta(\boldsymbol{x},t)-\eta^{\text{true}}(\boldsymbol{x},t)\right)^2}^t\right)^{1/2}}{H_s}, 
\end{equation}
where $\overline{(\cdot)}^t$ denotes a time average taken over a window where the error $\mathcal{E}^{g}$ is stationary, in this case $500T_p<t<1700T_p$. The spatial error $\mathcal{E}$ is plotted in figure \ref{fig:pointwise averaged}$(a)$. We see a low-error region near the body location, with $\mathcal{E}\leq 1\%$, and that the error increases up to $3\%$ as the distance from the body becomes larger. This is in general because of the higher correlation of the body motion to the nearby wave field. The low-error region near the body is of size close to the peak wave length, and extends longer in the $x$ direction due to the effect of advection, i.e., the information of wave field at up- and downstream of the body is better correlated with the body motion as it aligns with the major wave propagation direction. 

\begin{figure}
    \centering  \includegraphics[width=0.9\textwidth]{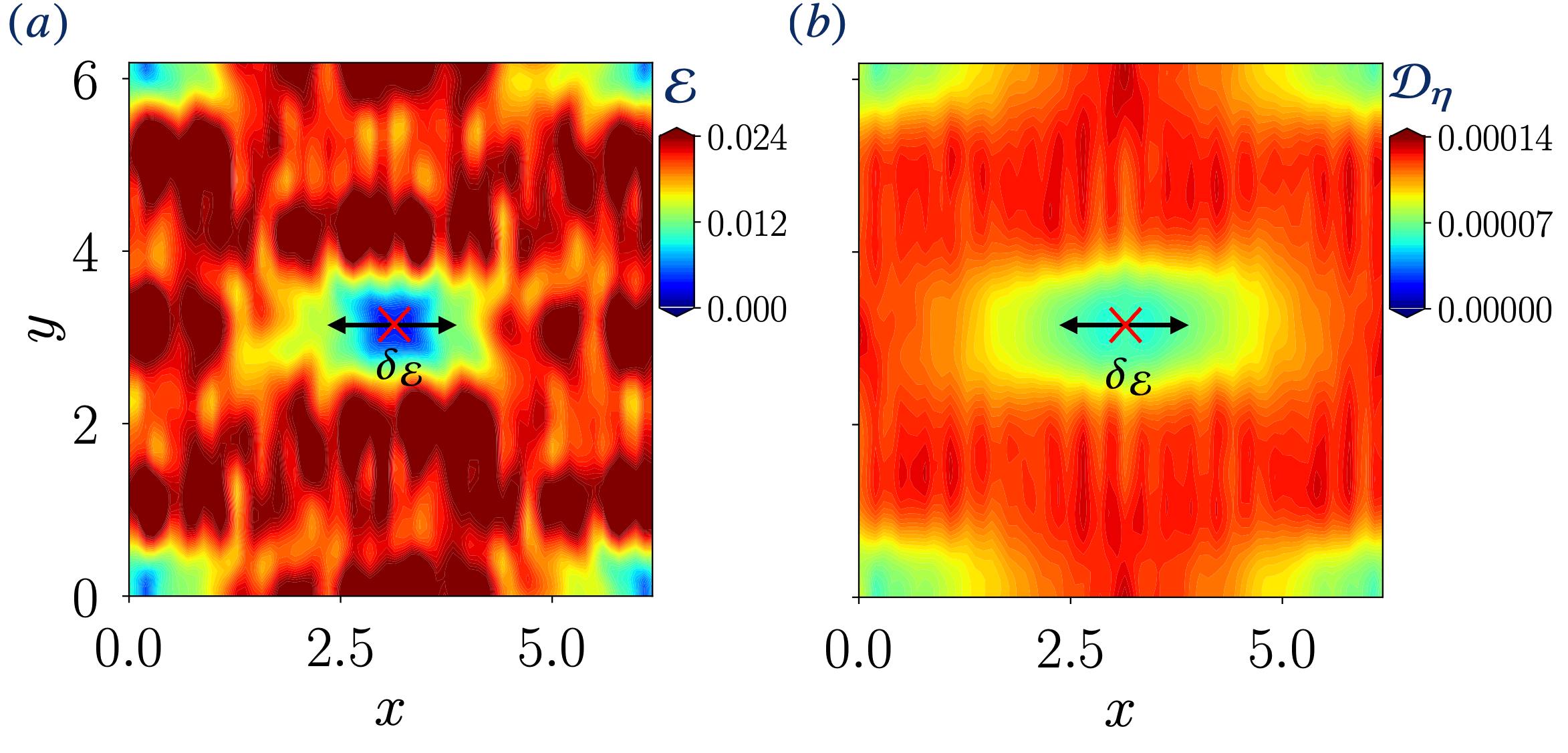}
    \caption{(a) $\mathcal{E}$ and (b) $\mathcal{D}_{\eta}$ for the reconstructed wave field, with the body marked (\textcolor{red}{$\times$}) and low-error regions size $\delta_{\mathcal{E}}$ denoted.}
    \label{fig:pointwise averaged}
\end{figure}

\begin{figure}
    \centering  \includegraphics[width=0.65\textwidth]{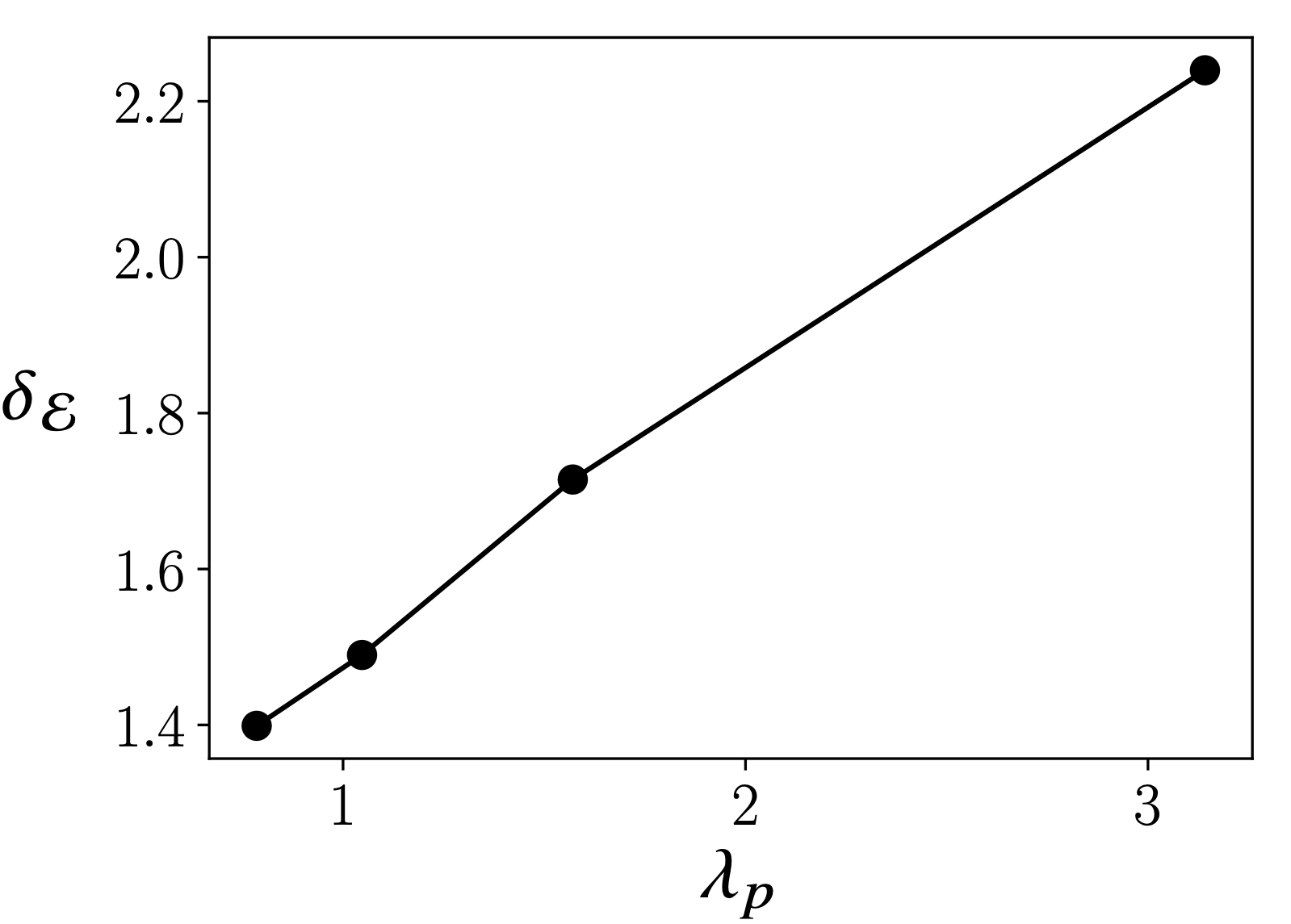}
    \caption{Dependence of $\delta_{\mathcal{E}}$ on $\lambda_p$. The four markers correspond to $\lambda_p = \{\frac{\pi}{4},\frac{\pi}{3},\frac{\pi}{2},\pi\}$. }
    \label{fig: low error region size}
\end{figure}

We note that in real applications, $\eta^{\text{true}}$ is not available, thus the error defined by (\ref{eq: error space}) cannot be obtained. However, the characteristics of the spatial error pattern can be alternatively constructed by evaluating the ensemble standard deviation, defined by
\begin{equation}\label{eq: ensemble cov eta}
    \mathcal{V}_{\eta}(\boldsymbol{x},t) = \frac{\sqrt{\overline{\left(\eta(\boldsymbol{x},t)-\overline{\eta}(\boldsymbol{x},t)\right)^2}}}{H_s}, 
\end{equation}
where $\overline{(\cdot)}$ denotes the ensemble average taken after each assimilation step. We further define a time-averaged ensemble standard deviation $\mathcal{D}_{\eta}(\boldsymbol{x})=\overline{\mathcal{V}_{\eta}(\boldsymbol{x},t)}^t$ and plot $\mathcal{D}_{\eta}$ in figure \ref{fig:pointwise averaged}$(b)$. We see a close resemblance of the pattern between \ref{fig:pointwise averaged}$(a)$ and $(b)$, although their absolute values differ. In principle, the quantity $\mathcal{D}_\eta(\boldsymbol{x})$ is a part of the epistemic uncertainty \citep{smith2024uncertainty}, that is, the uncertainty due to the state being under-constrained by the data. In this case, the wave state is only well constrained near the body, leading to a small $\mathcal{D}_\eta(\boldsymbol{x})$ there only. The similarity of spatial distribution between $\mathcal{D}_\eta(\boldsymbol{x})$ and $\mathcal{E}_\eta(\boldsymbol{x})$ indicates that the reconstruction uncertainty $\mathcal{D}_\eta(\boldsymbol{x})$ serves as a useful indicator of error distribution in applications where no knowledge of the true solution is available. 

The low-error region around the body corresponds to where the effective infusion of measurement information occurs. The size of this low-error region represents how far the measurement information can reach, and is shown to depend on the peak wavelength $\lambda_p=2\pi/k_p$. We define the low error region to have error $\mathcal{E}(\boldsymbol{x})$ lower than $30\%$ of the maximum value, and $\delta_{\mathcal{E}}$ to be the length of this region in $x$ direction through the body (see $\delta_{\mathcal{E}}$ marked in figure \ref{fig:pointwise averaged}). The values of $\delta_{\mathcal{E}}$ are computed from a series of assimilations with $\lambda_p = \{\frac{\pi}{4},\frac{\pi}{3},\frac{\pi}{2},\pi\}$ and are visualized in figure \ref{fig: low error region size}. A clear increase of $\delta_{\mathcal{E}}$ is observed with the increase of peak wavenumber. This is because the body motion correlates mostly with the wave motion inside one peak wavelength. Outside this range, such correlation decay, and the information provided by the body becomes less instructive in the assimilation. Therefore, the peak wavelength $\lambda_p$ serves as a characteristic lengthscale to describe the infusion process of body information into the wave field. 

\begin{figure}
    \centering  \includegraphics[width=1.0\textwidth]{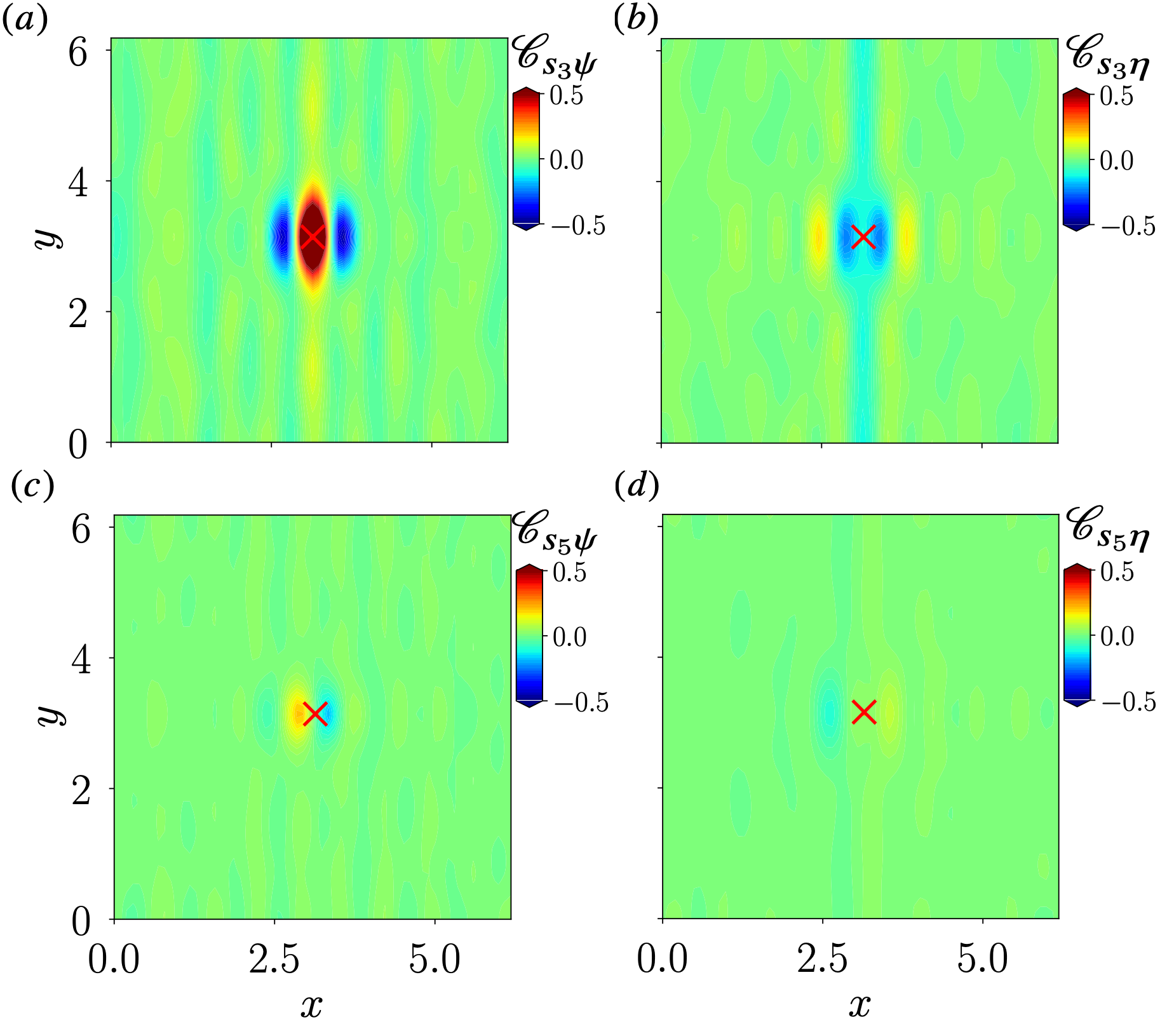}
    \caption{Ensemble correlations $(a)$ $\mathscr{C}_{s_3\psi}$, $(b)$ $\mathscr{C}_{s_3\eta}$, $(c)$ $\mathscr{C}_{s_5\psi}$ and $(d)$ $\mathscr{C}_{s_5\eta}$, with the body location marked (\textcolor{red}{$\times$}). }
    \label{fig:ensemble correlation eta s3}
\end{figure}

To further physically understand the high accuracy of wave reconstruction near the body, we study the motion-wave cross-correlation, which is the key to infer wave field from the body motion. For this purpose we define the ensemble correlation as
\begin{equation}\label{eq: ensemble cov eta}
    \mathscr{C}_{s^l\eta}(\boldsymbol{x},t) = \frac{\overline{\left(s^l(t)-\overline{s^l}(t)\right)\left(\eta(\boldsymbol{x},t)-\overline{\eta}(\boldsymbol{x},t)\right)}}
    {\sqrt{\overline{\left(s^l(t)-\overline{s^l}(t)\right)^2}}\sqrt{\overline{\left(\eta(\boldsymbol{x},t)-\overline{\eta}(\boldsymbol{x},t)\right)^2}}}. 
\end{equation}
The ensemble correlation $\mathscr{C}_{s^l\eta}$ represents how perturbations of $s^l$ and $\eta$ are correlated around the instantaneous ensemble mean state, relying on which EnKF assimilation is performed. A similar definition of $\mathscr{C}_{s^l\psi}$ can be reached by replacing $\eta$ with $\psi$ in (\ref{eq: ensemble cov eta}). Our goal is to show that high correlation region of $\mathscr{C}_{s^l\eta}$ and $\mathscr{C}_{s^l\psi}$ covers the high-accuracy region of wave reconstruction. Figure \ref{fig:ensemble correlation eta s3} shows four selected correlations under this principle, namely $\mathscr{C}_{s_3\psi}$, $\mathscr{C}_{s_3\eta}$, $\mathscr{C}_{s_5\psi}$ and $\mathscr{C}_{s_5\eta}$. In (a) and (b), we see that $s_3$ correlates well with $\psi$ around the body location and $\eta$ at up- and downstream of the body. This is clearly due to the phase lag between $\eta$ and $\psi$ of about $\pi/2$ along the wave profile, and that $s_3$ has a closer phase to $\psi$ at the body location under the current Cummins model. 
In (c) and (d), the correlations regarding $s_5$ exhibit weaker patterns, but the higher-correlation region in general occurs near the body. 
Furthermore, comparing (b) and (d), we see that the high-correlation regions at up- and downstream of the body exhibit different patterns, symmetric in (b) and anti-symmetric in (d). These different patterns suggest that different features of the surface are reconstructed from different motions in the correction associated with the Kalman filter (\ref{eq: analysis}). In this case, the symmetric correction in $\eta$ is provided by $s_3$ and anti-symmetric correction is provided by $s_5$. 

Following the above discussion, it is the collective effect of all 6-DoF measured motions that result in the low error in surface reconstruction seen in figure \ref{fig:pointwise averaged}. We further demonstrate this point by showing results with less number of DoFs measured. Figure \ref{fig:error history NDoF} shows the global error $\mathcal{E}^g$ including additional three cases with $\mathcal{M}= \{3\}$, $\mathcal{M}= \{2,3\}$ and $\mathcal{M}= \{1,2,3\}$.
The reconstruction accuracy drops slightly when the number of measured DoFs is reduced from $6$ to $2$, but deteriorates sharply when only the heave motion is measured ($\mathcal{M}=\{3\}$). This is because measuring heave alone makes the inverse problem poorly conditioned: wave fields propagating in different directions can produce similar heave responses. Therefore, heave alone does not provide sufficient information for accurate reconstruction of the wave profile.
\rv{In comparison, the measurement set $M=\{1,2,3\}$ already contains enough information to reconstruct the wave field. Specifically, $s_1$ and $s_2$ are sensitive to pressure gradient (and therefore surface elevation changes) in $x$- and $y$-directions, respectively; $s_3$ is sensitive to the surface elevation itself. Therefore, these three quantities already contain enough information to reconstruct the local wave field around the body}. 
The detailed effect of each DoF motion on the wave reconstruction can be a rich topic to be explored in future.

\begin{figure}
    \centering  \includegraphics[width=0.7\textwidth]{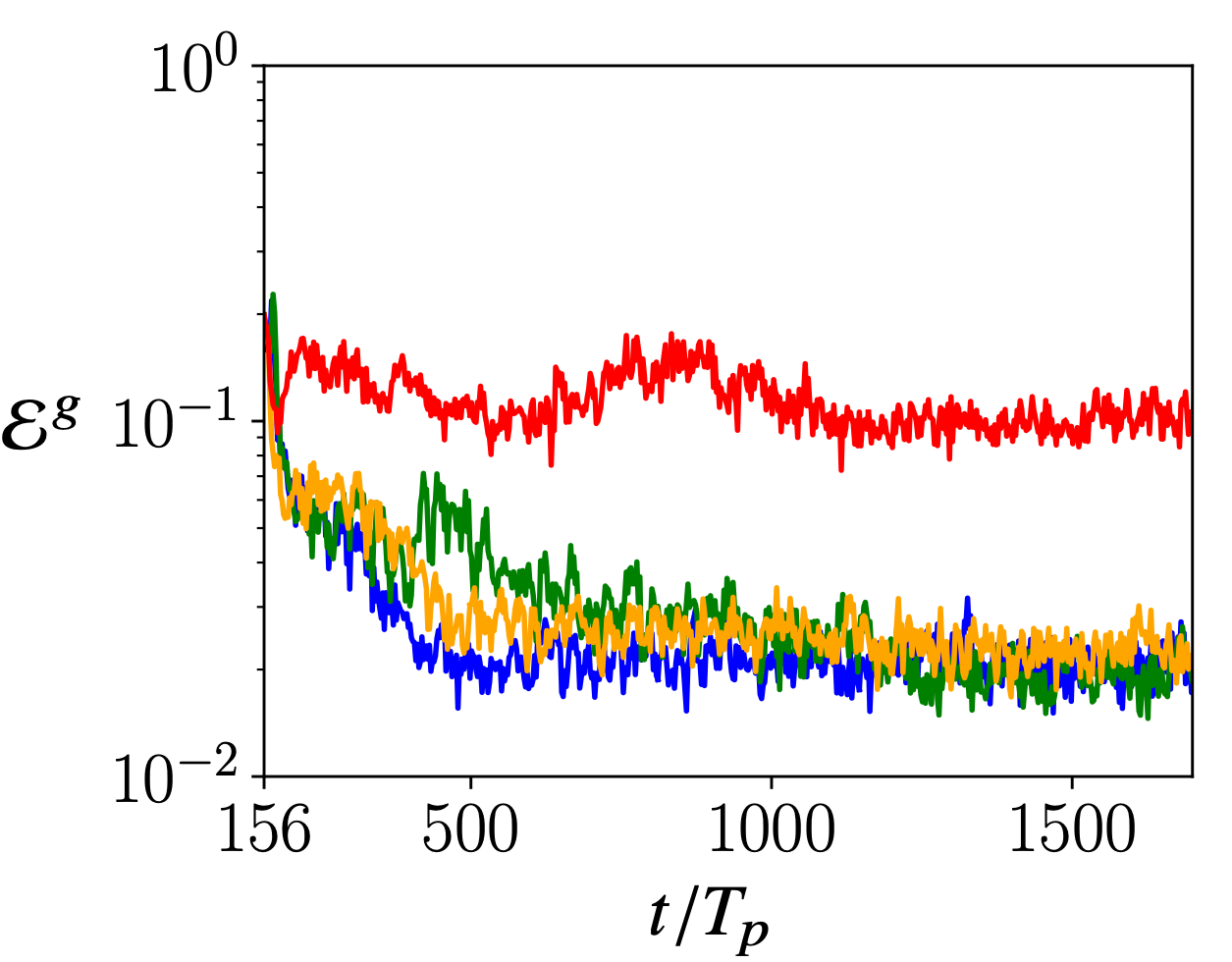}
    \caption{The global error $\mathcal{E}^g$ as a function of time, with measured DoFs  $\mathcal{M}=\{3\}$ (\fullred), $\{2,3\}$ (\fullorange), $\{1,2,3\}$ (\fullgreen) and  $\{1,2,3,4,5,6\}$(\full)} 
    \label{fig:error history NDoF}
\end{figure}

\rv{Apart from the measurement set $\mathcal{M}$, the sea state can also influence the reconstruction quality. As an example, we consider different significant wave heights $H_s=\{0.57\mathrm{m},5.7\mathrm{m}, 11.4\mathrm{m}\}$, which correspond to wave steepness $\epsilon=\{0.012,0.06,0.12\}$. For the two cases with larger $H_s$, the amplitudes of floating body motions are large, and the linearized assumption of the Cummins model no longer accurate. Nonetheless, the purpose of these cases is to show the effect of wave nonlinearity rather than body motion nonlinearity on the assimilation accuracy. Figure \ref{fig:error history steepness} shows the comparison of the global errors $\mathcal{E}^g$ as functions of time. The errors $\mathcal{E}^g(t)$ of all three cases reduce with time and saturate below $1\%$. The slight difference between the saturated error levels for different cases could be contributed by a lot of factors, including the nonlinear coupling between wave modes and the evolution of the energy spectrum. 
This set of tests demonstrates that our framework is effective for higher level of steepness (and nonlinearity).  }

\begin{figure}
    \centering  \includegraphics[width=0.7\textwidth]{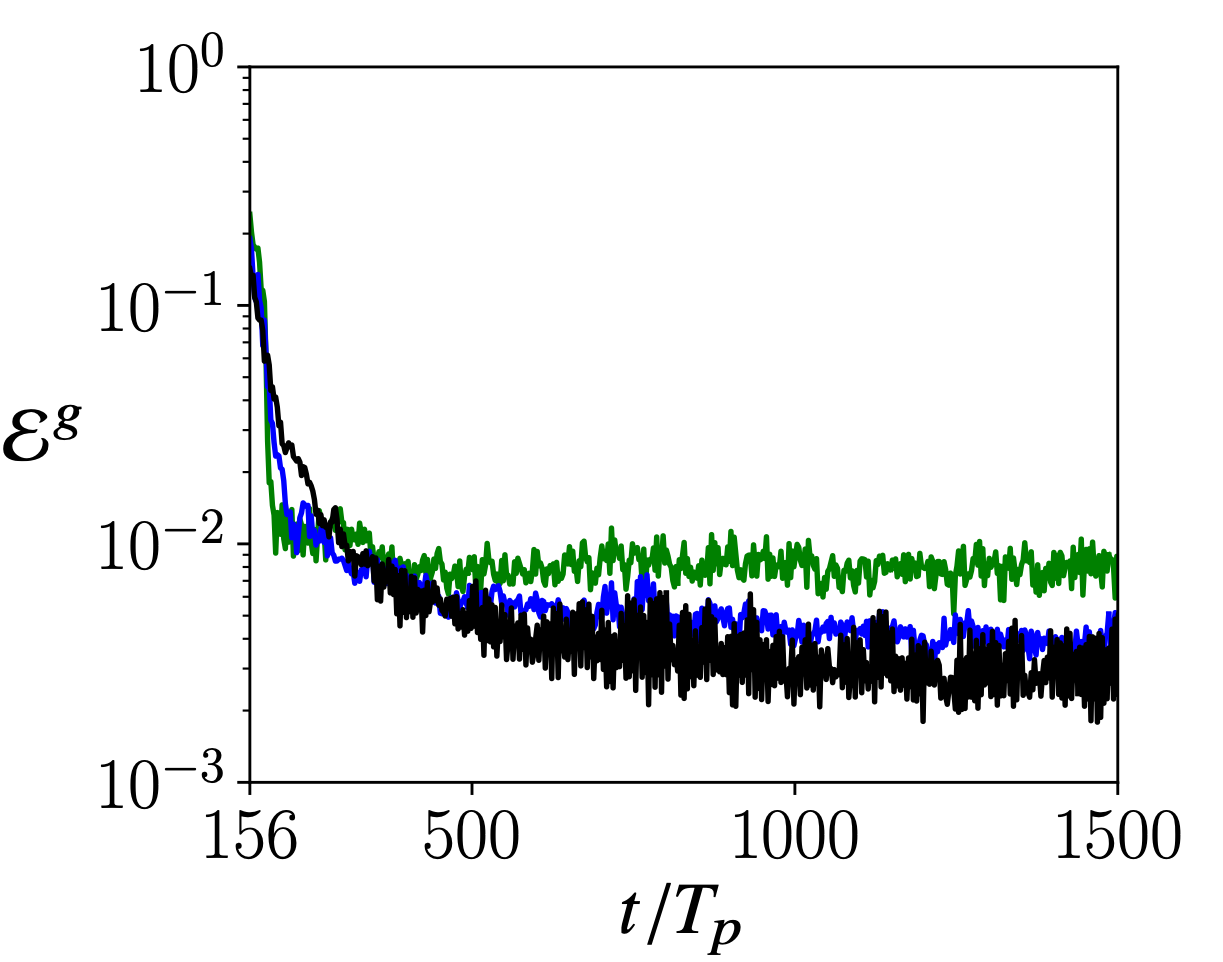}
    \caption{\rv{The global error $\mathcal{E}^g$ as a function of time, with $\epsilon=0.012$ (\fullgreen), $0.06$ (\full) and $0.12$ (\fullblack).}} 
    \label{fig:error history steepness}
\end{figure}

\subsection{Multiple floating bodies}\label{subsec: multiple ships}
\begin{figure}
    \centering  \includegraphics[width=0.7\textwidth]{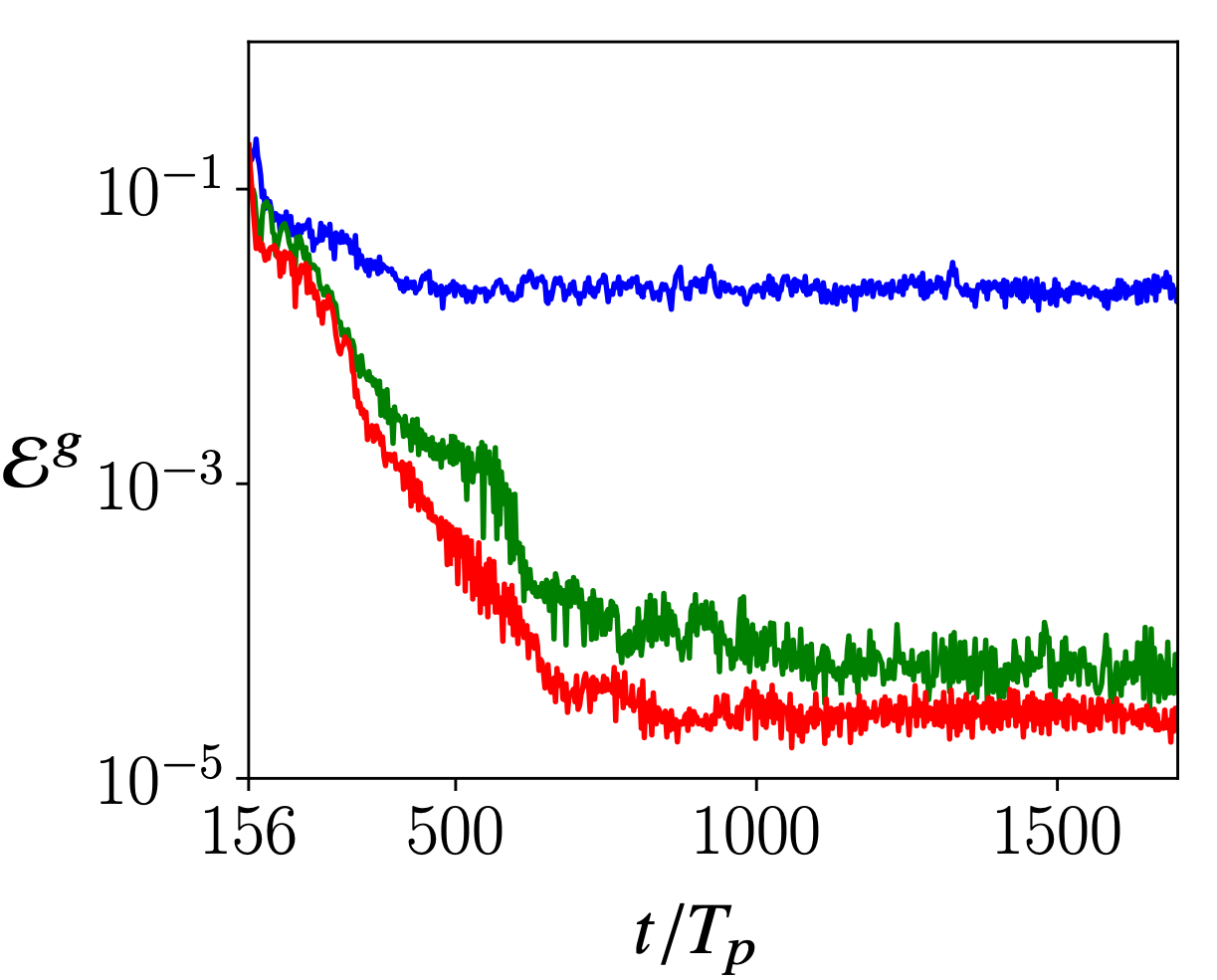}
    \caption{The global error $\mathcal{E}^g$ as a function of time, with measurements from one (\full{}), two (\fullgreen) and four (\fullred) floating bodies. }
    \label{fig:error time history 1, 2, 4 ships}
\end{figure}
\begin{figure}
    \centering  \includegraphics[width=1.0\textwidth]{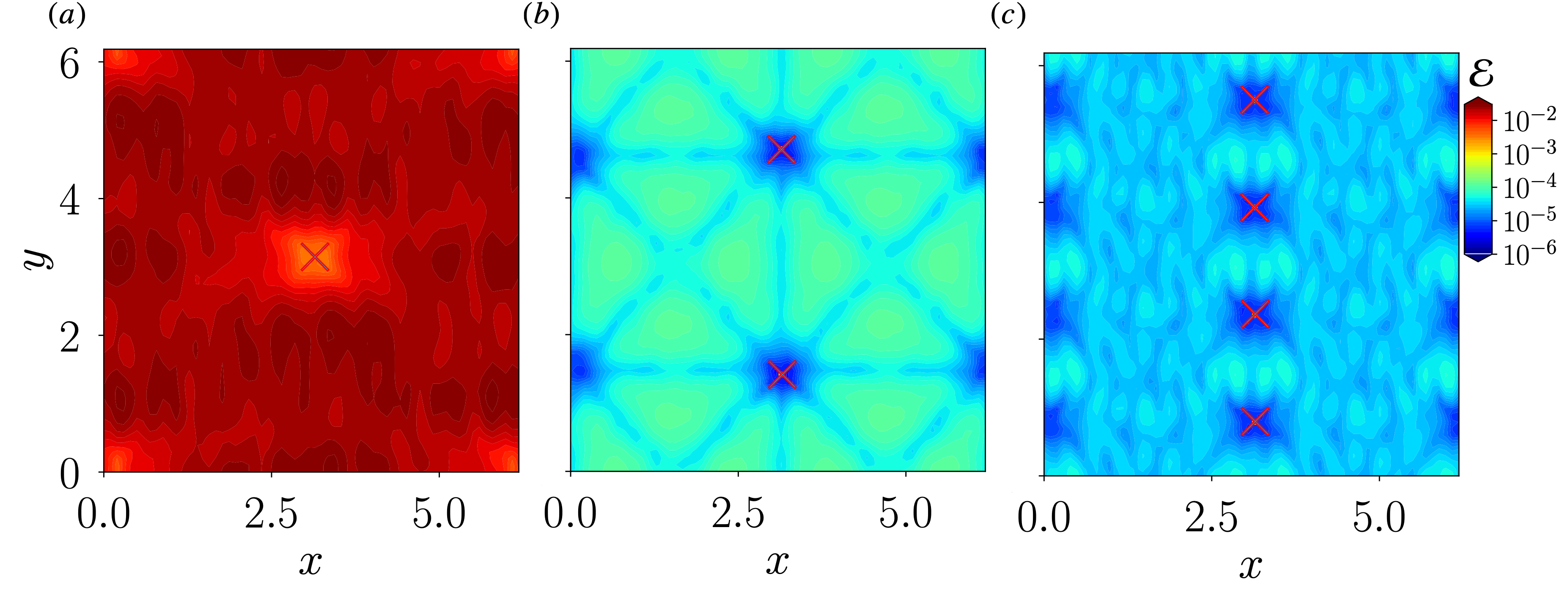}
    \caption{Comparison of $\mathcal{E}$ with measurements from $(a)$ one, $(b)$ two and $(c)$ four floating bodies at locations marked by (\textcolor{red}{$\times$}). }
    \label{fig:pointwise averaged 1 2 4 ships}
\end{figure}


\rv{In some applications including a ship fleet or an energy farm with floating turbine platforms, the motion measurements are often available from multiple floating bodies operating in the same period and spatial area \citep{mounet2022simultaneous,pujari2024real,mounet2023deriving}.} In this section, we show results with multiple floating bodies in two configurations: (i) two bodies and (ii) four bodies, each evenly distributed along the line $x=\pi$ (see figure \ref{fig:pointwise averaged 1 2 4 ships}). \rv{Although the same body geometry and RAO are assumed here, this is not in general necessary for our framework to be applicable}. In figure \ref{fig:error time history 1, 2, 4 ships}, we compare the time series of global error $\mathcal{E}^g$ obtained in these two cases to that in the single-body baseline case. We see that incorporation of two bodies reduces the saturated value of $\mathcal{E}^g$ by more than two orders of magnitude relative to the single-body case. Further increasing the number of bodies to four yields a more modest reduction of error of less than one order of magnitude. 

In general, the decrease of reconstruction error by incorporating more bodies comes from two factors. First, the effective reconstruction zone becomes larger as more bodies are included. Second, the measurements from different bodies act constructively in the inference of waves in the domain. Both points are demonstrated in figure \ref{fig:pointwise averaged 1 2 4 ships} which shows the spatial error $\mathcal{E}$ in the cases of 1, 2 and 4 floating bodies. We see an overall decrease of spatial error $\mathcal{E}$ as the body number increases. Moreover, there is also a significant decrease of $\mathcal{E}$ around the body location (in particular, the lowest $\mathcal{E}$ decreases from $3.3\times10^{-3}$ in (a), to $4.4\times 10^{-6}$ in (b) and to $4.0\times 10^{-6}$ in (c)). The decrease in reconstruction error is most significant when the body number increases from one to two, and becomes more marginal for the increase from two to four. The diminished return of the latter can be attributed to the growing correlation among measurements from different bodies, which introduces redundancy and limits the amount of independent information added to the EnKF update. These results highlight the value of incorporating multiple bodies, while also demonstrating the diminishing returns of dense and correlated observations.   

\subsection{Uncertain parameters}\label{subsec: results uncertain parameters}
\begin{figure}
    \centering  \includegraphics[width=0.8\textwidth]{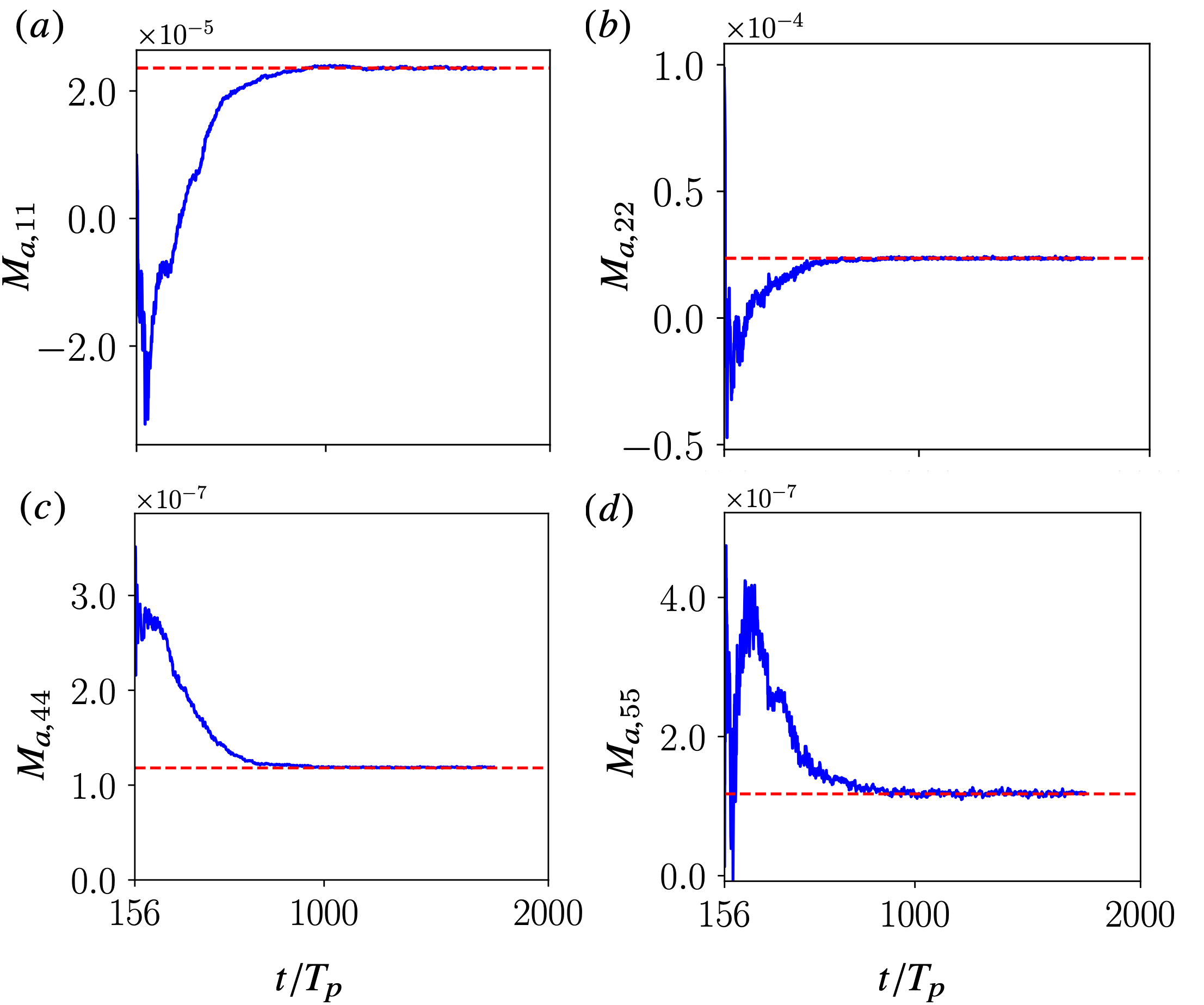}
    \caption{Comparison of $(a)$ $M_{a,11}$, $(b)$ $M_{a,22}$, $(c)$ $M_{a,44}$ and $(d)$ $M_{a,55}$ between the values from online estimation (\full) and the true value (\dashedred).} 
    \label{fig:M11 22 44 55 history}
\end{figure}

\begin{figure}
    \centering  \includegraphics[width=0.7\textwidth]{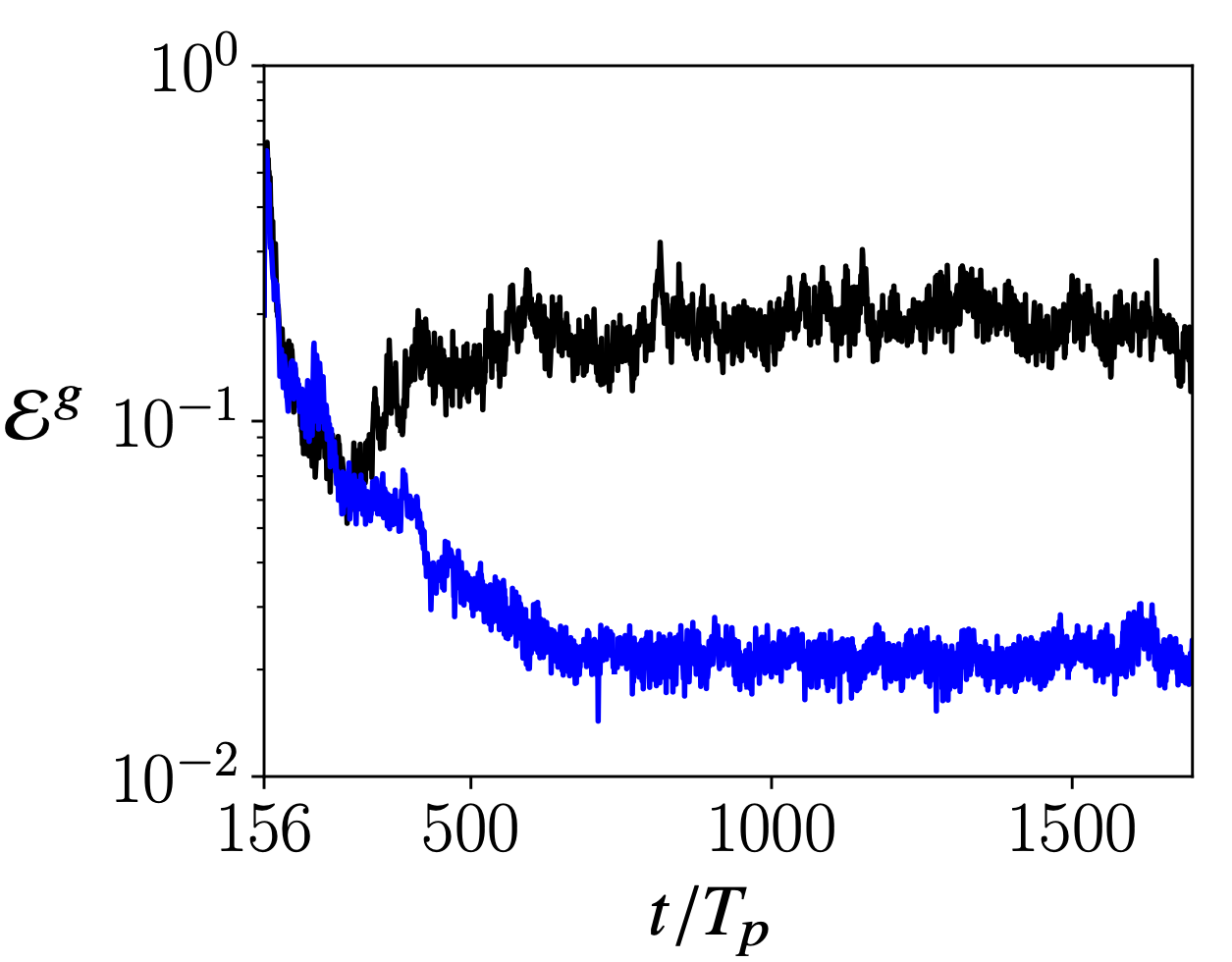}
    \caption{The global error $\mathcal{E}^g$ as a function of time, with parameter $\boldsymbol{M}_a$ from  online estimation (\full{}) and initial guess (\fullblack).   }
    \label{fig:Mopt error history}
\end{figure}

\begin{figure}
    \centering  \includegraphics[width=1.0\textwidth]{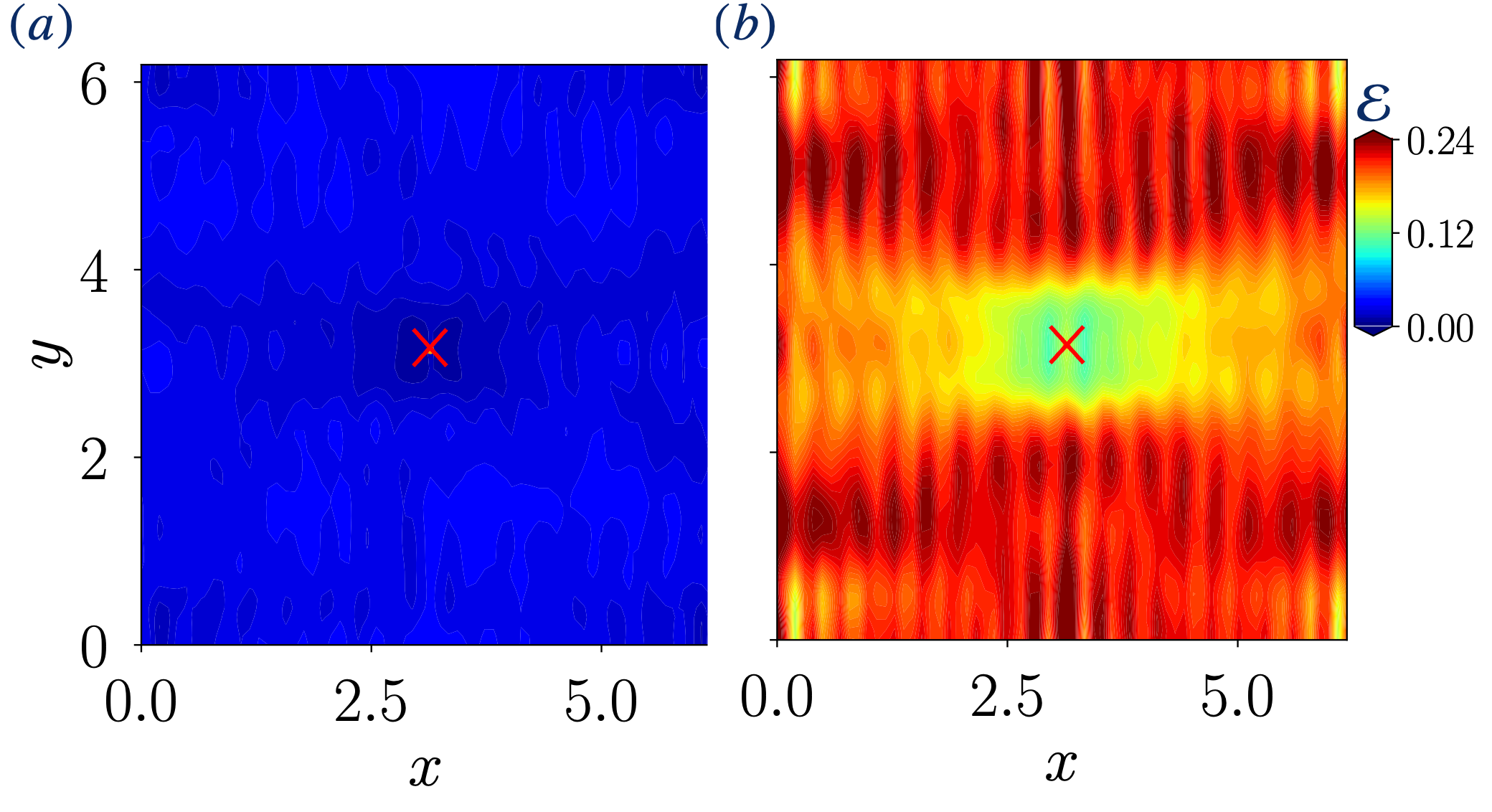}
    \caption{The spatial profile of $\mathcal{E}$ with parameters $\boldsymbol{M}_a$ from $(a)$ online estimation and $(b)$ initial guess, with the body location marked (\textcolor{red}{$\times$}).    }
    \label{fig:pointwise averaged opt vs noopt}
\end{figure}

We consider in this section a case with four of the diagonal elements of the added mass matrix $\boldsymbol{M}_a$ to have inaccurate initial estimations. In particular, the initial values of $M_{a,11}$, $M_{a,22}$, $M_{a,44}$, $M_{a,55}$ are set up by multiplying their true values (presented in Appendix \ref{sec: appendix Cummins}) with a random factor uniformly distributed in the range of $(0.4,4)$. We note that the real uncertainty associated with the hydrodynamic coefficients of a ship may be more complicated, very likely involving other added mass coefficients and wave-damping coefficients. The case we present here therefore represents a simplified setting to verify our data assimilation framework.

Figure \ref{fig:M11 22 44 55 history} shows the estimated values of the four uncertain added mass coefficients as a function of time. After an initial transient period, the estimated parameter values converge to the vicinity of their true values, which indicates the effectiveness of the maximum likelihood method presented in \S \ref{subsec: uncertain parameters}. \del{A closer scrutiny of figure \ref{fig:M11 22 44 55 history} suggests that the convergence of $M_{a,11}$ and $M_{a,55}$ seems slightly better than that of $M_{a,22}$ and $M_{a,44}$. This is probably because $M_{a,11}$ and $M_{a,55}$ are associated with $s^1$ and $s^5$ motions that are more sensitive to $x$-direction gradient of the wave profile, while $M_{a,22}$ and $M_{a,44}$, associated with $s^2$ and $s^4$, are affected more by the $y$-direction gradient. Since the incoming waves are relatively long-crested, $M_{a,11}$ and $M_{a,55}$ are better constrained by the measured data (as a way to think about this, any change in $M_{a,11}$ or $M_{a,55}$ results in more significant changes in $s^1$ and $s^5$ that are directly measured).  }

We finally present the reconstruction error with and without parameter estimation, where in the latter case the parameters retain their initial inaccurate values. Figure \ref{fig:Mopt error history} shows the error $\mathcal{E}^g$ as a function of time, where we see that $\mathcal{E}^g$ in the case with parameter estimation drops to 2\%, in contrast to 20\% in the case without parameter estimation. Figure \ref{fig:pointwise averaged opt vs noopt} shows the spatial error $\mathcal{E}$ for the two cases. In both cases, the spatial errors share a similar pattern, but the case with parameter estimation reach much lower error level overall. These results reinforce the necessity of online parameter estimation for ships in real sea, as a key for accurate reconstruction of the wave field.

\section{Conclusions} \label{sec:conclusions}
We have developed a new data assimilation framework to address the “ship-as-a-buoy” problem, for the purpose of obtaining a phase-resolved wave field from measured motion time series of floating bodies. The first step of the framework involves a RAO-based initial estimation of the wave field, which is used as an initial condition for the second step combining HOS and Cummins equation as dynamic models and EnKF as the data assimilation interface. In addition to application with one floating body, our approach can be naturally extended to cases with multiple floating bodies and cases where online estimation of model parameters is necessary. We provide verification of our framework in all above scenarios to showcase the accurate reconstruction, with a global error of $O(0.7\%)$ of the significant wave height that further improves with the increase of body numbers. The favorable reconstruction capability is further discussed in the context of ensemble correlations between the motion and the wave field.

\rv{Our work introduces new insight into the area of sea state estimation from ship motions. The ship motion measurement contains not only the spectral information of the wave (as those exploited by \citet{iseki2000bayesian,nielsen2006estimations}), but also phase information that is less commonly utilized. We demonstrate that it is possible to decode both parts of information collectively using EnKF data assimilation, and reconstruct the spatiotemporal wave field with high accuracy. Furthermore, we analyze the high-accuracy region near the body using the the ensemble correlation between the wave field and the body motions, and identify the peak wavelength as the characteristic lengthscale of assimilation. }

\rv{There are clear limitations in this work that should be addressed in the future. First, the floating body model adopted in this work is simplified and does not resolve a lot of complex mechanisms in ship motions. 
In real applications where floating bodies and ships have size similar to the wavelength, all of the physical mechanism involved in extreme ship motions should be adequately resolved in the floating body model. 
Second, the RAO-based initialization in the current framework is only accurate when the spreading angle is small. A systematic way of incorporating more wave traveling directions (in addition to the dominant wave direction) must be introduced to improve the initialization accuracy for larger spreading angle. 
Third, the discovery of the optimal strategy to incorporate the first and second points into the current framework would require extensive tests and trials, such that the overall method can be both efficient and accurate. 
}

This work opens up several possibilities for future research. Some immediate next tasks also include the optimal placement of multiple bodies and detailed understanding of the role of each DoF motion in reconstruction, as well as consideration of ship forward speed and path in the framework. In longer term, we are in the process of formulating the wave reconstruction problem from the perspective of observability theory, which will provide us theoretical bounds on the reconstruction to guide future applications.    

\rv{
\backsection[Data availability statement]
{The code and data used to produce the results presented in this study are
publicly available at \url{https://github.com/YDuME/EnKF_Ship_3D}.}
}

\backsection[Funding]
{The authors acknowledge financial support from the Office of Naval Research (N000142412535). }

\backsection[Declaration of interests]
{The authors report no conflict of interest.}

\appendix

\section{Analytical gradient of the RAO-based loss}\label{sec:RAO_loss_gradient}

To derive the analytical gradient of the loss function (\ref{eq:RAO loss}), we first introduce the vector variables
\begin{equation}
\hat{\mathbf{s}}_{\mathrm{m},\omega}
=
\begin{bmatrix}
\hat{s}^{\ell_1}_{\mathrm{m},\omega}\\
\vdots\\
\hat{s}^{\ell_{|\mathcal{M}|}}_{\mathrm{m},\omega}
\end{bmatrix}\in\mathbb{C}^{|\mathcal{M}|},
\qquad
\mathbf{Z}_{\mathrm{m}}(k,\rv{\theta(k)})
=
\begin{bmatrix}
Z^{\ell_1}(k,\rv{\theta(k)})\\
\vdots\\
Z^{\ell_{|\mathcal{M}|}}(k,\rv{\theta(k)})
\end{bmatrix}\in\mathbb{C}^{|\mathcal{M}|},
\end{equation} 
where $\mathcal{M}=\{\ell_1,\ldots,\ell_{|\mathcal{M}|}\}$. With these definitions, (\ref{eq:RAO loss}) is re-written as
\begin{equation}\label{eq:RAO_loss_vec_form_appendix}
\mathcal{J}_{\omega(k)}(\rv{a(k)},\rv{\theta(k)})
=
\frac{1}{2}
\left(\hat{\mathbf{s}}_{\text{m},\omega}-\rv{a(k)}\mathbf{Z}_{\mathrm{m}}(k,\rv{\theta(k)})\right)^\dagger
\left(\hat{\mathbf{s}}_{\text{m},\omega}-\rv{a(k)}\mathbf{Z}_{\mathrm{m}}(k,\rv{\theta(k)})\right), 
\end{equation}
where $(\cdot)^{\dagger}$ represents a conjugate transpose. Since $\mathcal{J}_{\omega(k)}$ depends on the complex variable $\rv{a(k)}$, Wirtinger derivatives are used, and $\rv{a(k)}$ and $\rv{a(k)}^\ast$ are treated as independent variables. Differentiation with respect to $\rv{a(k)}^\ast$ gives
\begin{equation}\label{eq:RAO_grad_wirtinger_ast_appendix}
\frac{\partial \mathcal{J}_{\omega(k)}}{\partial \rv{a(k)}^\ast}
=
-\frac{1}{2}\mathbf{Z}_{\mathrm{m}}(k,\rv{\theta(k)})^\dagger
\left(\hat{\mathbf{s}}_{\text{m},\omega}-\rv{a(k)}\mathbf{Z}_{\mathrm{m}}(k,\rv{\theta(k)})\right),
\end{equation}
and, equivalently,
\begin{equation}\label{eq:RAO_grad_wirtinger_appendix}
\frac{\partial \mathcal{J}_{\omega(k)}}{\partial \rv{a(k)}}
=
-\frac{1}{2}
\left(\hat{\mathbf{s}}_{\text{m},\omega}-\rv{a(k)}\mathbf{Z}_{\mathrm{m}}(k,\rv{\theta(k)})\right)^\dagger
\mathbf{Z}_{\mathrm{m}}(k,\rv{\theta(k)}).
\end{equation}

Differentiation of (\ref{eq:RAO_loss_vec_form_appendix}) with respect to the real variable $\rv{\theta(k)}$ gives
\begin{equation}\label{eq:RAO_grad_theta_component_appendix}
\frac{\partial \mathcal{J}_{\omega(k)}}{\partial \rv{\theta(k)}}
=
-\Re\left[
\rv{a(k)}^\ast
\left(\frac{\partial \mathbf{Z}_{\mathrm{m}}(k,\rv{\theta(k)})}{\partial \rv{\theta(k)}}\right)^{\dagger}
\left(\hat{\mathbf{s}}_{\text{m},\omega}-\mathbf{Z}_{\mathrm{m}}(k,\rv{\theta(k)})\rv{a(k)}\right)
\right].
\end{equation}

A closed expression for $\partial \mathbf{Z}_{\mathrm{m}}/\partial\rv{\theta(k)}$ follows from the frequency-domain RAO representation. We define short-hand notations
\begin{equation}\label{eq:RAO_matrix_vector_defs_appendix}
\mathbf{H}(\omega)=
-\omega^2(\mathbf{M}+\mathbf{A}(\omega))
+i\omega\mathbf{B}(\omega)+\mathbf{C}^R,
\end{equation}
and
\begin{equation}\label{eq:RAO_excitation_vector_def_appendix}
\mathbf{g}(k,\rv{\theta(k)})
=
\int_{A_w}
\begin{pmatrix}
\boldsymbol{n}\\
\boldsymbol{r}\times\boldsymbol{n}
\end{pmatrix}
\rho g\,e^{kz-i\boldsymbol{k}\cdot\boldsymbol{x}}\,
\mathrm{d}A(\boldsymbol{x},z),
\end{equation}
so that
\begin{equation}\label{eq:RAO_compact_Qg_appendix}
\mathbf{Z}_{\mathrm{m}}(k,\rv{\theta(k)})=\mathbf{P}\mathbf{H}(\omega)^{-1}\mathbf{g}(k,\rv{\theta(k)}). 
\end{equation}
The projection matrix $\mathbf{P}$ selects the measured DoFs from all six DoFs. For a fixed $\omega$ (hence fixed $k$), $\mathbf{H}(\omega)$ is independent of $\rv{\theta(k)}$, and
\begin{equation}\label{eq:RAO_dZ_dtheta_appendix}
\frac{\partial \mathbf{Z}_{\mathrm{m}}(k,\rv{\theta(k)})}{\partial \rv{\theta(k)}}
=
\mathbf{P}\mathbf{H}(\omega)^{-1}
\frac{\partial \mathbf{g}(k,\rv{\theta(k)})}{\partial \rv{\theta(k)}}.
\end{equation}
Substitution of (\ref{eq:RAO_compact_Qg_appendix}-\ref{eq:RAO_dZ_dtheta_appendix}) into (\ref{eq:RAO_grad_theta_component_appendix}) yields, 
\begin{equation}\label{eq:RAO_grad_theta_final_appendix}
\frac{\partial \mathcal{J}_{\omega(k)}}{\partial \rv{\theta(k)}}
=
-\Re\left[
\rv{a(k)}^\ast
\left(
\mathbf{P}\mathbf{H}(\omega)^{-1}
\frac{\partial \mathbf{g}(k,\rv{\theta(k)})}{\partial \rv{\theta(k)}}
\right)^\dagger
\left(\hat{\mathbf{s}}_{\text{m},\omega}-\rv{a(k)}\mathbf{P}\mathbf{H}(\omega)^{-1}\mathbf{g}(k,\rv{\theta(k)})\right)\right], 
\end{equation}
with $\partial \mathbf{g}/\partial\rv{\theta(k)}$ given by, 
\begin{equation}\label{eq:RAO_gtheta_explicit_appendix}
\frac{\partial \mathbf{g}(k,\rv{\theta(k)})}{\partial \rv{\theta(k)}}
=
\int_{A_w}
\begin{pmatrix}
\boldsymbol{n}\\
\boldsymbol{r}\times\boldsymbol{n}
\end{pmatrix}
\rho g\,e^{kz-i\boldsymbol{k}\cdot\boldsymbol{x}}
\left[-i\,k(-x\sin\rv{\theta(k)}+y\cos\rv{\theta(k)})\right]
\,\mathrm{d}A(\boldsymbol{x},z).
\end{equation}
Equations (\ref{eq:RAO_grad_wirtinger_ast_appendix}) and (\ref{eq:RAO_grad_theta_final_appendix}) provide the analytical gradient of (\ref{eq:RAO loss}) used by the L-BFGS optimizer.

\section{Parameters in the Cummins equation}\label{sec: appendix Cummins}

\begin{table}[htbp]
\centering
\caption{Value of the elements in the total mass matrix $\boldsymbol{M}+\boldsymbol{M}_a$. Units of the values vary by block: $\text{kg}$ for $i,j=1,2,3$, $\text{kg}\cdot \text{m}$ for $i=1,2,3$ and $j=4,5,6$ (or vice versa), and $\text{kg}\cdot \text{m}^2$ for $i,j=4,5,6$.
}
\label{tab:mass_matrix_cummins}
\renewcommand{\arraystretch}{1.15}
\setlength{\tabcolsep}{6pt}
\small

\begin{tabular}{@{} c c
                *{6}{>{\centering\arraybackslash}p{1.4cm}} @{}}
\toprule
& & \multicolumn{6}{c}{\textbf{Columns}} \\
\cmidrule(lr){3-8}
& & $s^1$ (surge) &
    $s^2$ (sway) &
    $s^3$ (heave) &
    $s^4$ (roll) &
    $s^5$ (pitch) &
    $s^6$ (yaw) \\
\midrule
\multirow{6}{*}{\textbf{Rows}} &
$s^1$ (surge) & $153514$ & $0$ & $0$ & $0$ & $109587$ & $0$ \\
& $s^2$ (sway)  & $0$ & $153514$ & $0$ & $-109587$ & $0$ & $0$ \\
& $s^3$ (heave) & $0$ & $0$ & $279194$ & $0$ & $0$ & $0$ \\
& $s^4$ (roll)  & $0$ & $-109587$ & $0$ & $2009531$ & $0$ & $0$ \\
& $s^5$ (pitch) & $109587$ & $0$ & $0$ & $0$ & $2009543$ & $0$ \\
& $s^6$ (yaw)   & $0$ & $0$ & $0$ & $0$ & $0$ & $1341721$ \\
\bottomrule
\end{tabular}

\vspace{4pt}
\begin{minipage}{0.92\linewidth}
\footnotesize
\end{minipage}
\end{table}

We report the values of the parameters appearing in the Cummins equation (\ref{eq:Cummins}), namely the total mass matrix $\boldsymbol{M}+\boldsymbol{M}_a$, the restoring matrix $\boldsymbol{C}^R$, and the memory kernel $\boldsymbol{K}(t)$. These quantities are obtained from the online WEC-SIM project repository. The total mass matrix consists of four blocks: $(M+M_a)_{1:3,1:3}$, corresponding to translation; $(M+M_a)_{1:3,4:6}=(M+M_a)_{4:6,1:3}^{\top}$, corresponding to translation-rotation coupling; and $(M+M_a)_{4:6,4:6}$, corresponding to rotation. These blocks carry different physical units, and the matrix entries are reported in SI units in table \ref{tab:mass_matrix_cummins}.

\begin{table}[htbp]
\centering
\caption{Value of the elements in the restoring matrix $\boldsymbol{C}_R$. Units of the values vary by block: $\text{kg}\cdot \text{s}^{-2}$ for $i,j=1,2,3$, $\text{kg}\cdot \text{m}\cdot \text{s}^{-2}$ for $i=1,2,3$ and $j=4,5,6$ (or vice versa), and $\text{kg}\cdot \text{m}^2\cdot \text{s}^{-2}$ for $i,j=4,5,6$.
}
\label{tab:restoring_matrix_cummins}
\renewcommand{\arraystretch}{1.15}
\setlength{\tabcolsep}{6pt}
\small

\begin{tabular}{@{} c c
                *{6}{>{\centering\arraybackslash}p{1.4cm}} @{}}
\toprule
& & \multicolumn{6}{c}{\textbf{Columns}} \\
\cmidrule(lr){3-8}
& & $s^1$ (surge) &
    $s^2$ (sway) &
    $s^3$ (heave) &
    $s^4$ (roll) &
    $s^5$ (pitch) &
    $s^6$ (yaw) \\
\midrule
\multirow{6}{*}{\textbf{Rows}} &
$s^1$ (surge) & $57$ & $0$ & $0$ & $0$ & $0$ & $0$ \\
& $s^2$ (sway)  & $0$ & $57$ & $0$ & $0$ & $0$ & $0$ \\
& $s^3$ (heave) & $0$ & $0$ & $763311$ & $0$ & $0$ & $0$ \\
& $s^4$ (roll)  & $0$ & $0$ & $0$ & $6056889$ & $0$ & $0$ \\
& $s^5$ (pitch) & $0$ & $0$ & $0$ & $0$ & $6098393$ & $0$ \\
& $s^6$ (yaw)   & $0$ & $0$ & $0$ & $0$ & $0$ & $562$ \\
\bottomrule
\end{tabular}

\vspace{4pt}
\begin{minipage}{0.92\linewidth}
\footnotesize
\end{minipage}
\end{table}

The entries of $\boldsymbol{C}^R$ are computed analytically for the oblate spheroid, based on a horizontal waterplane located at the body center. Owing to the symmetry of the body, only three entries of $\boldsymbol{C}^R$ are non-zero:
\begin{equation}
C^R_{3,3} = \rho g A_{wp}, \qquad
C^R_{4,4} = \rho g \left(I_x + V(z_B-z_G)\right), \qquad
C^R_{5,5} = \rho g \left(I_y + V(z_B-z_G)\right).
\end{equation}
Here, $A_{wp}=\pi a_x a_y$ is the waterplane area, $V=\frac{2}{3}\pi a_x a_y a_z$ is the immersed volume, and $z_B=-\frac{3}{8}a_z$ and $z_G=-2$ denote the vertical coordinates of the centres of buoyancy and gravity relative to the waterplane, respectively. The quantities $I_x$ and $I_y$ are the second moments of the waterplane area, given by
\begin{equation}
I_x = \int_{A_{wp}} x^2 \,\mathrm{d}A = \frac{1}{4}\pi a_x^3 a_y, \qquad
I_y = \int_{A_{wp}} y^2 \,\mathrm{d}A = \frac{1}{4}\pi a_x a_y^3.
\end{equation}

To improve the numerical stability of the $s^1$, $s^2$, and $s^6$ components of the Cummins equation, we introduce three artificial mooring coefficients, $C^R_{11}$, $C^R_{22}$, and $C^R_{66}$, computed as \citep{kim2008technical}
\begin{equation}
T_i = 2\pi \sqrt{\frac{M_{ii}+M_{a,ii}}{c_{ii}}}, \qquad
C^R_{ii} = \left(M_{ii}+M_{a,ii}\right)\left(\frac{2\pi}{T_i}\right)^2,
\qquad i=1,2,6.
\end{equation}
Here, $T_i$ an user-specific period, and is chosen to be much larger than the peak wave period so that the wave-induced body motion spectrum is not significantly altered by the artificial mooring. In the present work, we take $T_i = 300~\mathrm{s} \approx 30T_p$. The full matrix $\boldsymbol{C}^R$ is listed in table \ref{tab:restoring_matrix_cummins}.

\begin{figure}
    \centering
    \includegraphics[width=0.9\textwidth]{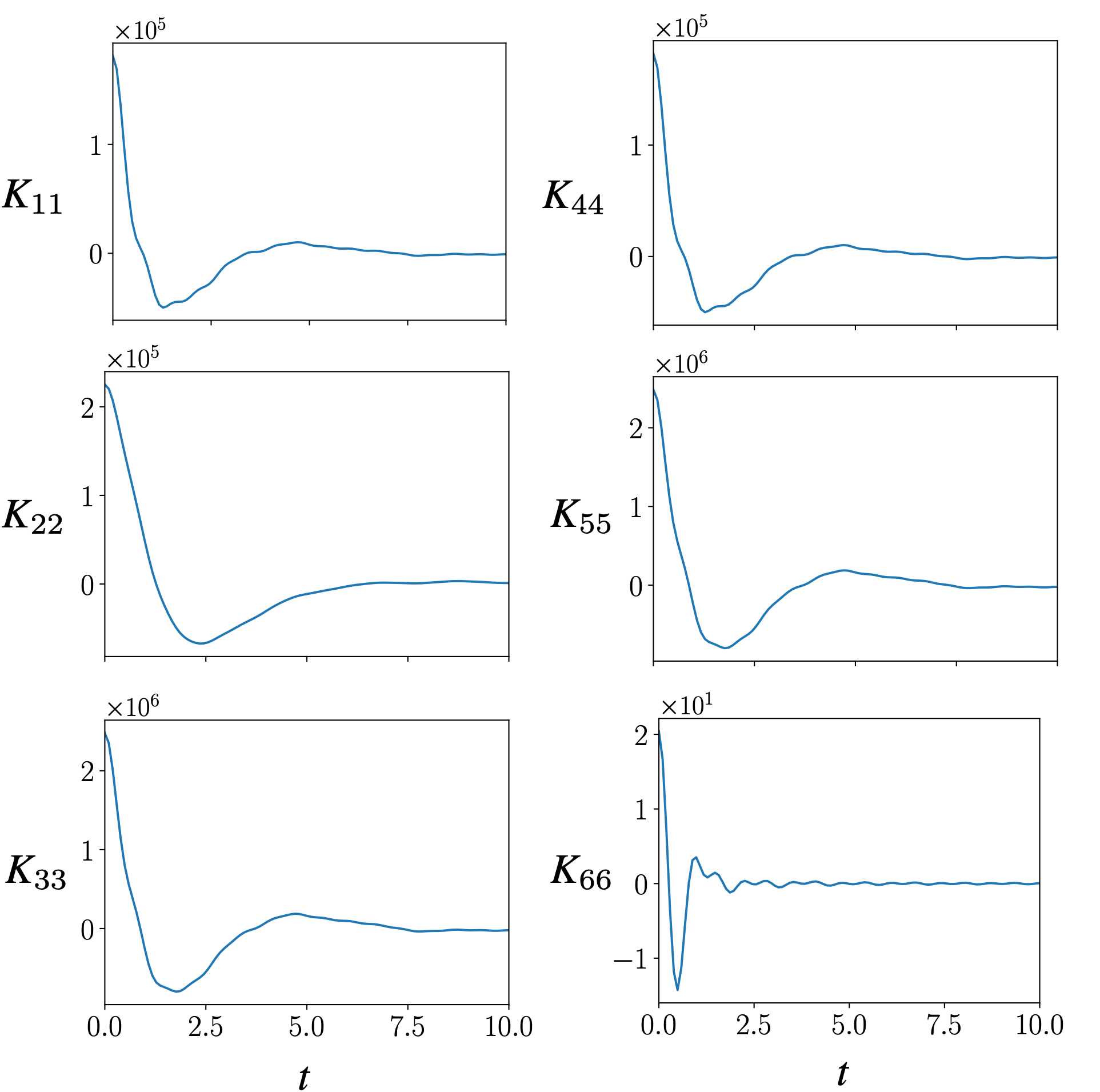}
    \caption{Time histories of the diagonal entries of the memory kernel $\boldsymbol{K}(t)$, plotted in standard units. }
    \label{fig: memory kernel}
\end{figure}

The memory kernel $\boldsymbol{K}(t)$ in the Cummins equation (\ref{eq:Cummins}) has 36 entries, each of which is a function of time. The diagonal entries are shown in figure \ref{fig: memory kernel}, while the remaining entries can be found in the online WEC-SIM project repository. A sample time history of the six DoF displacements and velocities is provided in figure \ref{fig: 6DoF s v sample}.

\begin{figure}
    \centering
    \includegraphics[width=0.9\textwidth]{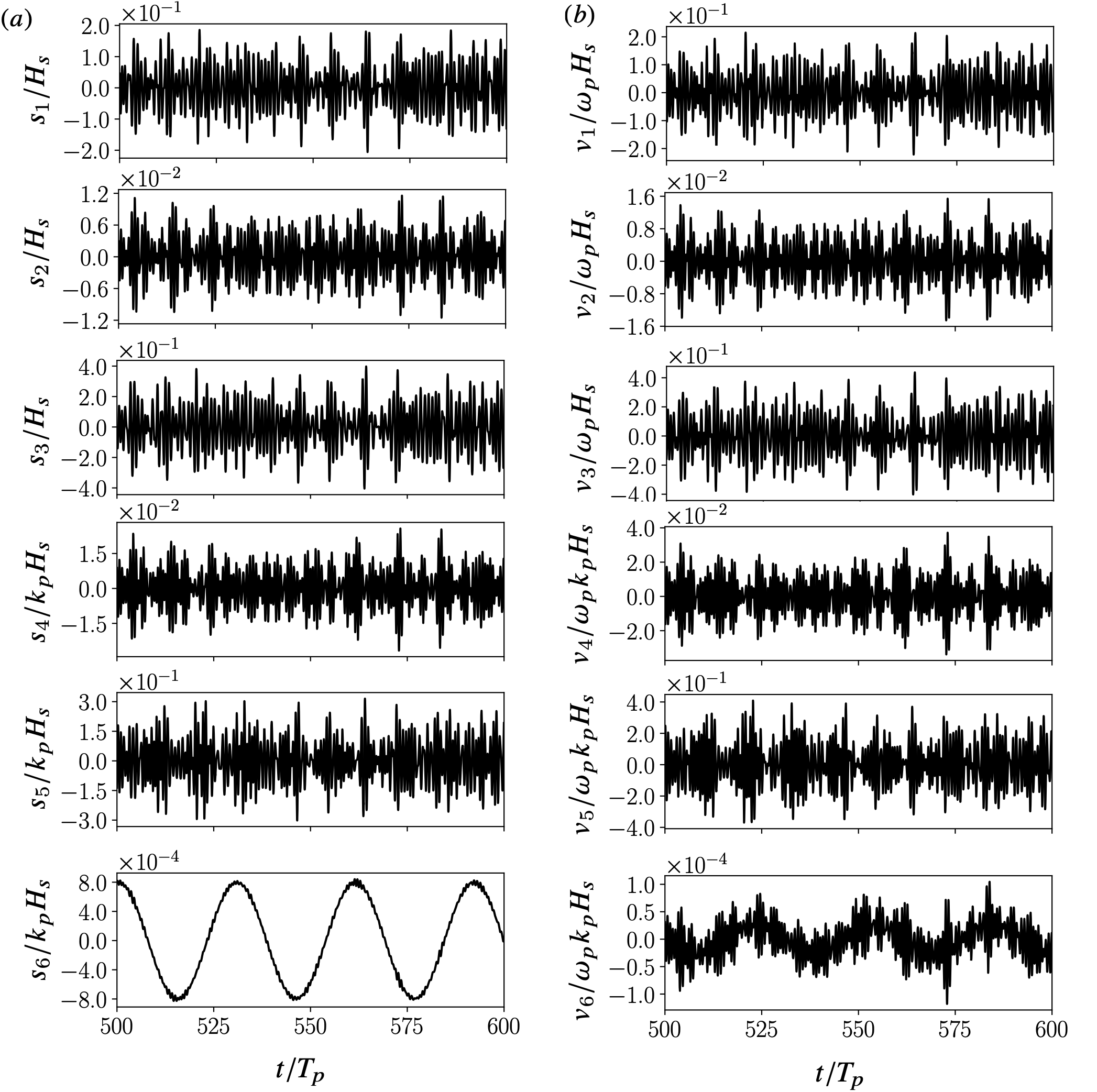}
    \caption{A sample time history of the six-DoF displacements $(a)$ and velocities $(b)$ of the floating body under the wave excitation. }
    \label{fig: 6DoF s v sample}
\end{figure}

\section{Detailed expression of likelihood functions}\label{sec: appendix EM}
The computation of the likelihood function $\tilde{L}$ in (\ref{eq: expectation}) rely on the conditional joint probability density $p(\boldsymbol{s}_{\text{m},j}, \mathbf{q}_{j} \mid \boldsymbol{\theta})$. This distribution $p(\boldsymbol{s}_{\text{m},j}, \mathbf{q}_{j} \mid \boldsymbol{\theta})$ is decomposed, by the law of total probability, into, 
\begin{equation}\label{eq: pdf decomposition}
      p(\boldsymbol{s}_{\text{m},j}, \mathbf{q}_{j} \mid \boldsymbol{\theta}) =p(\boldsymbol{s}_{\text{m},j} \mid \mathbf{q}_{j}, \boldsymbol{\theta}) p(\mathbf{q}_{j} \mid \boldsymbol{\theta}).   
\end{equation}
Assume the observation model
$\boldsymbol{s}_{\mathrm{m},j}=\mathbf H\mathbf q_j+\boldsymbol{\epsilon}_{\mathrm{m},j}$ with
$\boldsymbol{\epsilon}_{\mathrm{m},j}\sim\mathcal N(\mathbf 0,\mathbf R)$.
The conditional density $p(\boldsymbol{s}_{\mathrm{m},j}\mid \mathbf q_j,\boldsymbol{\theta})$ is therefore Gaussian and given by
\begin{equation}\label{eq: pdf observation}
p(\boldsymbol{s}_{\mathrm{m},j}\mid \mathbf q_j,\boldsymbol{\theta})
=
\frac{1}{\sqrt{(2\pi)^{|\mathcal M|}\det(\mathbf R)}}
\exp\!\left(
-\frac12\left\Vert \boldsymbol{s}_{\mathrm{m},j}-\mathbf H\mathbf q_j\right\Vert^2_{\mathbf R^{-1}}
\right).
\end{equation}
Similarly, the density $p(\mathbf{q}_j \mid \boldsymbol{\theta})$ corresponds to numerical error or model inaccuracy, 
\begin{equation}\label{eq: pdf equation}
    p(\mathbf{q}_{j}\mid\boldsymbol{\theta}) =\frac{1}{\sqrt{(2\pi)^d \det{\mathbf{\Sigma}_p}}}\exp\left(-\frac{1}{2}\left\Vert\mathbf{q}_{j}-\mathcal{N}_\theta(\mathbf{q}_{j-1},,\dot{\boldsymbol{s}}_{P}))\right\Vert^2_{\mathbf{\Sigma}_p^{-1}}\right), 
\end{equation}
where $d=2N+12$ is the dimension of the state vector,  and $\mathbf{\Sigma}_p$ is the covariance of error of the physical dynamics of the wave-body interaction. The expressions (\ref{eq: pdf decomposition}-\ref{eq: pdf equation}) can be plugged into equation (\ref{eq: expectation}) for numerical evaluation. 

We further note that the parameter $\boldsymbol{\theta}$ is irrelevant to the measurement process and does not appear in the right hand side of (\ref{eq: pdf observation}). Therefore, the gradient $\nabla_{\boldsymbol{\theta}}\log(p(\boldsymbol{s}_{\text{m},j} \mid \mathbf{q}_{j}, \boldsymbol{\theta}))$ is zero, and the loss function $\tilde{L}(\boldsymbol{\theta}\mid\boldsymbol{\theta}^{(\iota)})$ is expressed as, 
\begin{equation}\label{eq: loss1}
    \tilde{L}(\boldsymbol{\theta}\mid\boldsymbol{\theta}^{(\iota)})=\mathbb{E}_{\mathbf{q}_{j}\mid \boldsymbol{s}_{\text{m},j},\boldsymbol{\theta}^{(\iota)}}\left[-\frac{1}{2}\left\Vert\mathbf{q}_{j}-\mathcal{N}_\theta(\mathbf{q}_{j-1},\dot{\boldsymbol{s}}_P)\right\Vert^2_{\mathbf{\Sigma}_p^{-1}}\right] + C_1, 
\end{equation}
where the constant $C_1$ does not depend on $\boldsymbol{\theta}$. In addition, in the context of estimating uncertain body parameters, the parameter $\boldsymbol{\theta}$ only appears in the Cummins equation, not in the ocean surface wave equations. Through a Monte-Carlo approximation over the ensemble members, the loss (\ref{eq: loss1}) is further simplified into, 
\begin{equation}\label{eq: loss2}
\tilde{L}(\boldsymbol{\theta}\mid\boldsymbol{\theta}^{(\iota)})
\approx\frac{1}{N_{\text{ens}}}\sum_{n=1}^{N_{\text{ens}}}\left[-\frac{1}{2}\left\Vert
    \begin{pmatrix}
\boldsymbol{s}^{\text{a},n}_{j} \\
\dot{\boldsymbol{s}}^{\text{a},n}_{j} 
\end{pmatrix}-\mathcal{N}_\theta^c(\boldsymbol{s}^{\text{a},n}_{j-1},\dot{\boldsymbol{s}}^{\text{a},n}_{j-1},\dot{\boldsymbol{s}}_{P})\right\Vert^2_{\mathbf{\Sigma}_p^{c,-1}}\right] + C_2, 
\end{equation}
where $\mathcal{N}_\theta^c$ denotes the solution mapping of the first-order Cummins equation (\ref{eq: Cummins first order}) from $t_{j-1}$ to $t_j$, $\mathbf{\Sigma}_p^{c}$ is the covariance matrix of the uncertainty contained in the Cummins equation, and the constant $C_2$ does not depend on $\boldsymbol{\theta}$. 

To evaluate $\nabla_{\boldsymbol{\theta}}\tilde{L}$ in (\ref{eq: loss2}), we differentiate the one-step solution operator $\mathcal{N}^c_\theta$ using an adjoint formulation of the first-order Cummins equation. For each ensemble member $n$, we take the endpoint analysis states $(\boldsymbol{s}^{\mathrm{a},n}_{j-1},\dot{\boldsymbol{s}}^{\mathrm{a},n}_{j-1})$ and $(\boldsymbol{s}^{\mathrm{a},n}_{j},\dot{\boldsymbol{s}}^{\mathrm{a},n}_{j})$, together with the memory effect $\dot{\boldsymbol{s}}_{P}$, as constants when computing derivatives. 
Consequently, we capture the sensitivity of $\tilde{L}$ on $\boldsymbol{\theta}$ only through the forward solution $(\boldsymbol{s}^n(t),\dot{\boldsymbol{s}}^n(t))$ on $t\in[t_{j-1},t_j]$, and neglect the dependence on $\boldsymbol{\theta}$ through $\boldsymbol{s}^{\mathrm{a},n}_{j-1},\dot{\boldsymbol{s}}^{\mathrm{a},n}_{j-1}, \dot{\boldsymbol{s}}_{P}$ during earlier time. 
Such a setting reduces the computational cost and complexity of the method, meanwhile produces accurate results (figure \ref{fig:M11 22 44 55 history}). We introduce adjoint variables $\boldsymbol{\lambda}^{n}_s(t)$ and $\boldsymbol{\lambda}^{n}_{\dot{s}}(t)$ associated with $(\boldsymbol{s}^n,\dot{\boldsymbol{s}}^n)$, which satisfy the adjoint Cummins equation:

\begin{subequations}\label{eq:adjoint_cummins_first_order}
\begin{align}
-\frac{\mathrm{d}\boldsymbol{\lambda}^{n}_s}{\mathrm{d}t}
+(\boldsymbol{C}^R)^\top \boldsymbol{\lambda}^{n}_{\dot{s}}
&=\boldsymbol{0}, \\
-\left(\boldsymbol{M}+\boldsymbol{M}_a\right)^\top \frac{\mathrm{d}\boldsymbol{\lambda}^{n}_{\dot{s}}}{\mathrm{d}t}(t)
-\boldsymbol{\lambda}^{n}_s(t)
+\int_t^{t_j} \boldsymbol{K}(\tau-t)^\top \boldsymbol{\lambda}^{n}_{\dot{s}}(\tau)\,\mathrm{d}\tau
&=\boldsymbol{0},
\end{align}
\end{subequations}
with terminal conditions
\begin{subequations}\label{eq:adjoint_cummins_terminal_cond}
\begin{align}
\begin{pmatrix}\boldsymbol{\lambda}_s^{n}(t_j)\\
(\boldsymbol{M}+\boldsymbol{M}_a)^{\top}\boldsymbol{\lambda}^{n}_{\dot{s}}(t_j)\end{pmatrix}&= 
{\mathbf{\Sigma}_p^{c,-1}}
\left(
    \begin{pmatrix}
\boldsymbol{s}^{\text{a},n}_{j} \\
\dot{\boldsymbol{s}}^{\text{a},n}_{j} 
\end{pmatrix}-\mathcal{N}_\theta^c(\boldsymbol{s}^{\text{a},n}_{j-1},\dot{\boldsymbol{s}}^{\text{a},n}_{j-1},\dot{\boldsymbol{s}}_{P})\right), 
\end{align}
\end{subequations}
The adjoint system (\ref{eq:adjoint_cummins_first_order}-\ref{eq:adjoint_cummins_terminal_cond}) is solved backward-in-time numerically using a discrete adjoint algorithm. The gradient of the loss $\tilde{L}$ with respect to the added-mass matrix
$\boldsymbol{\theta}=\boldsymbol{M}_a$ appear from the explicit
dependence of the Cummins equation on $\boldsymbol{M}_a$ through the inertial
term $(\boldsymbol{M}+\boldsymbol{M}_a)\,\mathrm{d}\dot{\boldsymbol{s}}^{n}/\mathrm{d}t$.
Using the adjoint variable $\boldsymbol{\lambda}^{n}_{\dot{s}}(t)$, the gradient $\nabla_{\boldsymbol{M}_a}\tilde{L}$ is obtained as, 
\begin{equation}\label{eq:grad_Ma_single_sample}
\nabla_{\boldsymbol{M}_a}\tilde{L}
=
\frac{1}{N_{\text{ens}}}\sum_{n=1}^{N_{\text{ens}}}\int_{t_{j-1}}^{t_j}
\boldsymbol{\lambda}^{n}_{\dot{s}}(t)
\left(\frac{\mathrm{d}\dot{\boldsymbol{s}}^{n}}{\mathrm{d}t}(t)\right)^\top
\,\mathrm{d}t.
\end{equation}

\rv{\section{A case with large $\beta$} \label{sec: appendix large beta}}

\begin{figure}
    \centering  \includegraphics[width=0.9\textwidth]{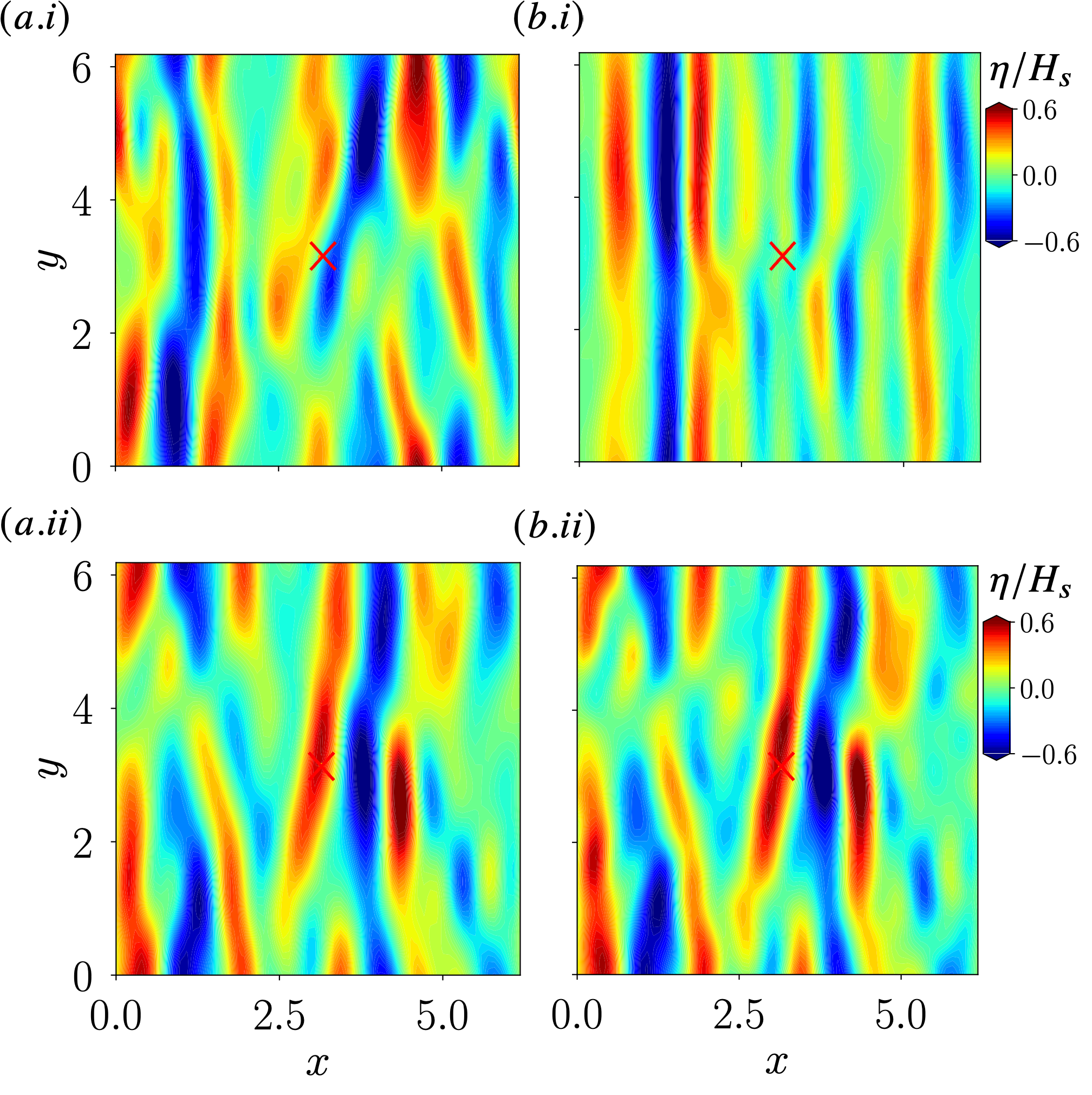}
    \caption{\rv{The surface elevation $\eta(\boldsymbol{x},t)$ for the true solution $(a)$ and assimilated solution $(b)$ at the beginning $(i)$ and the end $(ii)$ of the EnKF assimilation, with $\beta=\pi/3$}}
    \label{fig: spreading angle}
\end{figure}

\rv{Here we report a data assimilation case with $\beta=\pi/3$ and other parameters adopted from \S \ref{sec:results}, and compare the true solution (figure \ref{fig: spreading angle}$(a)$) with the assimilated solution (figure \ref{fig: spreading angle}$(b)$). Since the spreading angle is wide and the `one dominant wave' assumption is violated, the surface elevation $\eta$ from the RAO based initialization (figure \ref{fig: spreading angle}$(b.i)$) contains substantial error compared to that of true solution (figure \ref{fig: spreading angle}$(a.i)$). However, the RAO-based initialization is further improved by the EnKF data assimilation. At the end of the assimilation window, the assimilated surface elevation $\eta$ (figure \ref{fig: spreading angle}$(b.ii)$) is very close to the true elevation (figure \ref{fig: spreading angle}$(a.ii)$). In summary, although the RAO-based initialization deteriorate and contain large error for wave states with large spreading angle, the large error is reduced significantly by the EnKF data assimilation, and an estimation of instantaneous surface elevation $\eta$ is still produced largely consistent with the true $\eta$.}  

\bibliographystyle{jfm}
\bibliography{reference}

\end{document}